\documentclass[%
 reprint,
 amsmath,amssymb,
 aps,
]{revtex4-2}
\usepackage{graphicx}       
\usepackage{tikz}     
\usepackage{comment}
\usepackage{amsmath}        
\usetikzlibrary{quantikz2}  
\usepackage{pgfplots}       
\usepackage{graphicx}       
\usepackage{tabularx}       
\usepackage{algorithm}      
\usepackage{algpseudocode}  
\usepackage{subcaption}
\graphicspath{ {./image/} }
\def\highlightCircuitTopLabel#1#2#3#4{\gategroup[#1,steps=#2,style={dashed,rounded corners,fill=#3!20,inner xsep=2pt},background]{{\sc #4}}}
\def\highlightCircuitBottomLabel#1#2#3#4{\gategroup[#1,steps=#2,style={dashed,rounded corners,inner xsep=2pt,fill=#3!20},background,label style={label position=below,anchor=north,yshift=-0.2cm}]{{\sc #4}}}
\usepackage{graphicx}
\usepackage{dcolumn}
\usepackage{bm}
\usepackage{lipsum}
\usepackage{hyperref}

\begin{document}

\preprint{APS/123-QED}

\title{Robust Implementation of Discrete-time Quantum Walks in Any Finite-dimensional Quantum System}

\author{Biswayan Nandi$^{1}$}
\author{Sandipan Singha$^{1}$}
\author{Ankan Datta$^{2}$}
\author{Amit Saha$^{3,4}$}
\email{abamitsaha@gmail.com}
\author{Amlan Chakrabarti$^{1}$}

\affiliation{$^{1}$A. K. Choudhury School of Information Technology, University of Calcutta, India\\
$^{2}$Institute of Radiophysics and Electronics, University of Calcutta, Kolkata, India\\$^{3}$Institut National de Recherche en Informatique et en Automatique (Inria), Paris, France\\$^{4}$École Normale Supérieure (ENS), PSL University, Paris, France}

\date{\today}

\begin{abstract}
Research has shown that quantum walks can accelerate certain quantum algorithms and act as a universal paradigm for quantum processing. The discrete-time quantum walk (DTQW) model, owing to its discrete nature, stands out as one of the most suitable choices for circuit implementation. Nevertheless, most current implementations are characterized by extensive, multi-layered quantum circuits, leading to higher computational expenses and a notable decrease in the number of confidently executable time steps on current quantum computers. Since quantum computers are not scalable enough in this NISQ era, we also must confine ourselves to the ancilla-free frontier zone. Therefore, in this paper, we have successfully cut down the circuit cost concerning gate count and circuit depth by half through our proposed methodology in qubit systems as compared to the state-of-the-art increment-decrement approach. Furthermore, for the engineering excellence of our proposed approach, we implement DTQW in any finite-dimensional quantum system with akin efficiency. To ensure an efficient implementation of quantum walks without requiring ancilla, we have incorporated an intermediate qudit technique for decomposing multi-qubit gates. Experimental outcomes hold significance far beyond the realm of just a few time steps, laying the groundwork for dependable implementation and utilization on quantum computers.
\end{abstract}

\maketitle


\section{Introduction}
    The field of quantum computing \cite{feynman2018simulating, nielsenchuang} is in rapid development in the current era. Quantum computers are far more efficient in doing certain tasks than our best state-of-the-art classical computing algorithms. They can leverage quantum mechanical phenomena like superposition, entanglement, and measurement giving them a significant edge over classical computers. For example, Grover's search \cite{grover1996fast} can search in $O(\sqrt{N})$ time where $N$ is the size of the input space. In 1994, Peter Shor showed that we can prime factor a number much more efficiently using quantum computers \cite{shor1994algorithms}. Like these quantum algorithms, quantum walk algorithms have also applications in various fields such as quantum simulation, quantum search algorithms, and quantum cryptography. They also offer potential advantages over classical algorithms in certain scenarios, particularly in tasks that involve searching through large databases or graphs.
    
    Quantum walks \cite{aharonov1993quantum} is the quantum mechanical counterpart of classical random walks. Random walks have been used to simulate the motion of molecules in gas or liquid, also known as Brownian motion \cite{brownian1947}, unpredictable fluctuating stock prices \cite{stockprice1964}, PageRank \cite{pagerank} algorithms in search engines like Google, and recommendation algorithms also make use of it. Therefore, there is a significant amount of use cases for random walk algorithms but how is quantum walk different from classical random walk? We get a normal distribution of the probability of finding the walker in different states of the space. Therefore, around the state from which the walker starts walking has the highest probability, or in other words, the walker is most likely to return to where it started. In contrast, a quantum walker is the exact opposite. In a quantum walk, the probability of finding the walker around the state it started is the lowest. It is highest on the states farthest from the starting position. Therefore, we can say that, quantum walks move faster than their classical counterparts \cite{kempe2003quantum}, which can be useful for certain applications such as in developing cryptographic applications \cite{crypto2020}. 

    Two types of quantum walks are proposed in the literature: discrete-time quantum walks (DTQW) \cite{feynman2010quantum} and continuous-time quantum walks (CTQW) \cite{farhi1998quantum}. In the first model, we have a walker, a coin, and an evolution operator which is applied to both the walker and the coin in discrete time steps. On the other hand, in the second model, we have a walker and an evolution that can be applied to the walker with no timing restrictions. In this article, we have only worked with the first model i.e., DTQW. Before we discuss our motivation for implementing a scalable generalized quantum walk algorithm, we address why one should study quantum walks in the first place i.e., its applications. \textit{Childs} in his work showed that quantum walks can be considered a universal computational model \cite{childs2009universal, childs2013universal}. \textit{Ambainis} showed that quantum walks can be used to efficiently solve the element distinctness problem which is a problem of finding two distinct items among $N$ given ones \cite{ambainis2007quantum}. The application domain of quantum walks is not bounded by theoretical computer science only. We can use quantum walks to simulate natural processes like photosynthesis \cite{engel2007evidence}. The primary challenge in quantum computing lies in optimizing qubit utilization and circuit depth to maximize the implementation of Discrete-Time Quantum Walks (DTQW) within the constraints of qubit quantity and coherence time for its applications.

    \textbf{Motivation:} In the initial development of quantum walk algorithms \cite{kendon2006random} and in recent developments \cite{wadhia2024cycle} also, for a general quantum walk implementation, without considering special cases such as quantum walks on complete graphs \cite{douglas2009efficient}, the naive approach \cite{ambainis2001one}  is used in binary quantum systems. This is a generalized approach for implementing $n$-qubit quantum walks, albeit, this is inefficient concerning gate count and circuit depth. The noise level is also colossal because of its inefficient circuit structure. Our main motivation is to develop an efficient generalized approach to robust circuit implementation of DTQW that can also be scaled up to any finite-dimensional quantum system to make it noise-resilient. In \cite{singh2021quantum}, the authors have proposed a generalized quantum walk implementation for 5-qubit quantum hardware, but not scalable to the $n$-qubit systems. Another efficient quantum circuit for the DTQW on the cycle has been presented as a notable advancement in \cite{Razzoli_2024}. By employing just one QFT and one IQFT, the approach in \cite{Razzoli_2024} markedly enhances the most efficient state-of-the-art QFT-based implementation. Albeit, this QFT-based implementation is not efficient in this ancilla-free frontier zone \cite{Preskill_2018} and is also difficult to scale to any finite-dimensional quantum systems due to its complex circuit structure. In \cite{9410395, saha1}, the authors have shown an efficient implementation for $d$-dimensional quantum walks in qudit systems where $d$ is odd, but the circuit for even dimensions remains inefficient because it is still based on the increment-decrement approach due to engineering challenges. 
    
    \textbf{Contribution:} With the background of the implementation of DTQW, in this article, we have successfully cut down the circuit cost concerning gate count and circuit depth by half through our proposed methodology in qubit systems as compared to the naive increment-decrement approach. Certainly, this is the first-of-its-kind approach apart from naive increment-decrement, which is generalized for any $n$-qubit systems. Because of its engineering excellence, we also show that with slight modification, the proposed approach is generalized for any finite-dimensional quantum system. 
    
    
    Our key contributions are as follows:
    \begin{itemize}
        \item We optimize the circuit for DTQW by reducing the circuit depth and gate count. We have developed a generalized algorithm that is theoretically twice as efficient as the naive increment-decrement approach. From now on we will call it the \textbf{Enhanced Increment-Decrement} approach.
         \item Our proposed algorithm is generalized in the binary space, meaning it works for any number of qubits. However, our further aim is to extend this generalization to higher dimensions as well. Through our proposed approach, by setting the Least Significant Bit (LSB) as any even dimension when $d \ge 4$  and adjusting the dimensionality of the other Most Significant Bits (MSB) for any qudits, our proposed algorithm can be generalized for any $d$-dimensional space.

        \item We have further compared the naive approach and our proposed approach in terms of gate count, circuit depth, and effect of noise after Toffoli decomposition and we have observed that in every comparison, our proposed approach has outperformed the naive approach significantly.

    \end{itemize}

 In this paper, first, we discuss some basic concepts and gates in quantum computing in section \ref{background-section}. Then we discuss the mathematical foundation and implementation of the naive quantum walk approach in section \ref{naive-walk-approach}. Next, we explain our proposed methodology in binary systems in detail in section \ref{our-walk-approach} and compare our approach with the naive approach in section \ref{results}. We then discuss the implementation of our approach in higher dimensions in section \ref{our-walk-approach-higher-dimension}. Finally, we conclude our remarks in \ref{conclusion}.

\section{Background}\label{background-section}
In this section, our discussion will focus on some basics of quantum computing and quantum walks.

    \subsection{Quantum computing}
\begin{figure}[!h]
            \centering
            \scalebox{.5}{
            \begin{quantikz}
                \lstick{\ket{q_0}}&\ctrl{1}&\\
                \lstick{\ket{q_1}}&\ctrl{1}&\\
                \lstick{\ket{q_2}}&\targ{}&
            \end{quantikz}
            $\equiv$
            \begin{quantikz}
                &&&&\ctrl{2}&&&&\ctrl{2}&&\ctrl{1}&&\ctrl{1}&\gate{T}&&\\
                &&\ctrl{1}&&&&\ctrl{1}&&&&\targ{}&\gate{T^{\dag}}&\targ{}&\gate{T}&&\\
                &\gate{H}&\targ{}&\gate{T^{\dag}}&\targ{}&\gate{T}&\targ{}&\gate{T^{\dag}}&\targ{}&&&&&\gate{T}&\gate{H}&
            \end{quantikz}}
            \caption{2-controlled Toffoli decomposition using Clifford+T gate set}
            \label{fig:toffoli_decomposition_clifford}
        \end{figure}
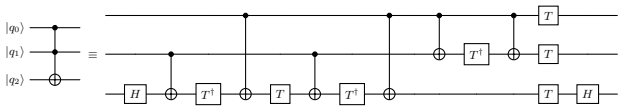

        The bit is the fundamental concept of classical computation and classical information. Quantum computation and quantum information are built upon an analogous concept, the quantum bit or qubit for short \cite{nielsenchuang}. As classical bits can be either $0$ or $1$, a qubit can be in a continuum of states between $0$ and $1$ until it is observed or measured which is called superposition. For example, a qubit can be in the probabilistic state $\alpha\ket{0}+\beta\ket{1}$, where $\alpha$ and $\beta$ can be complex numbers. Changes occurring in a quantum state can be described using the language of quantum computation. Analogous to the way a classical computer is built from an electrical circuit containing wires and logic gates, a quantum computer is built from a quantum circuit containing wires and elementary quantum gates to carry around and manipulate the quantum information \cite{nielsenchuang}. We have used the following gates for our proposed approach: Pauli X (NOT) Gate, Hadamard Gate, Controlled NOT (CX) Gate, Toffoli (CCX) Gate, and Multi-controlled Toffoli (C$^{n}$X) Gate and its generalized versions. A decomposition using the Clifford+T gate set of the Toffoli gate in binary systems is shown in Figure \ref{fig:toffoli_decomposition_clifford}, which is used to estimate the gate resources in this paper in binary quantum systems.

    \subsection{Discrete-time quantum walks (DTQW)} \label{naive-walk-approach}
    \subsubsection{Mathematical background of DTQW}
        As a mathematical illustration, DTQW is applied over 1D space which is an Integer line. A coin is tossed, and according to that, the walker shifts its position to the right or left. Let $H_c$ be the Hilbert space of the coin spanned by the basis set $\{\ket{\uparrow}, \ket{\downarrow}\}$ and $H_x$ be the Hilbert space of the position states spanned by the basis set $\{\ket{x}\}_{x\in \mathbb{Z}}$. The unitary operator for a single step of the quantum walk is given by
        \begin{equation}
            U=S(C\otimes I)
        \end{equation}
        where $C$ is the coin toss operator and $S$ is the shift operator that is applied to the walker after the toss. $\otimes$ is tensor product. In the case of a general coin,
        \begin{equation}
           \scalebox{.75}{  $C=cos\theta\ket{\uparrow}\bra{\uparrow}+e^{i\phi_1}sin\theta\ket{\uparrow}\bra{\downarrow}+e^{i\phi_2}sin\theta\ket{\downarrow}\bra{\uparrow}-e^{i(\phi_1+\phi_2)}cos\theta\ket{\downarrow}\bra{\downarrow}$}
        \end{equation}
        where $\theta\in[0, 2\pi)$ and $\phi_1, \phi_2\in[0, \pi)$. We refer to $\theta$ as rotation and $\phi_1, \phi_2$ as phase parameters of the general coin \cite{jayakody}. But here we have considered only the Haramard coin for simplicity. So having $\theta=\frac{\pi}{4}$ and $\phi_1, \phi_2=0$ we get,
        \begin{equation}
            C=\frac{1}{\sqrt{2}}\begin{bmatrix}
                1 & 1\\
                1 & -1
            \end{bmatrix}
        \end{equation}
        Finally, the unitary operator for shifting the walker is defined as \cite{saha1},
        \begin{equation}
            S=\sum_{x\in\mathbb{Z}}{(\ket{\uparrow}\bra{\uparrow}\otimes\ket{x+1}\bra{x}+\ket{\downarrow}\bra{\downarrow}\otimes\ket{x-1}\bra{x})}
        \end{equation}

    \subsubsection{Implementation of quantum walks in naive approach}
        The components for increment and decrement in a 3-qubit system in the naive implementation of DTQW are shown in Figure \ref{fig:incdeccomponents}. After a coin toss, which is a simple Hadamard transform in this case, we need to do both increment and decrement operations based on the state of the coin. The circuit for a single step of DTQW in this naive setting is shown in Figure \ref{fig:naivecircuit}. For $n$ qubits, we can have $2^{n-1}-1$ steps of DTQW. The position state mapping for the circuit shown in Figure \ref{fig:naivecircuit}, after it is applied 3 times is as Table \ref{tab:naive_mapping}.
        \begin{figure}[!h]
            \centering
             \scalebox{.6}{
            \begin{quantikz}
                \lstick{\ket{q_0}}&\ctrl{1}&\ctrl{1}&\targ{}&\\
                \lstick{\ket{q_1}}&\ctrl{1}&\targ{}&&\\
                \lstick{\ket{q_2}}&\targ{}& &&\\
                \lstick{\ket{coin}}&\ctrl{-1}&\ctrl{-2}&\ctrl{-3}&
            \end{quantikz}
            \hspace{2cm}
            \begin{quantikz}
                \lstick{\ket{q_0}}& &\targ{}&\ctrl{1}&\ctrl{1}&&\\
                \lstick{\ket{q_1}}& & &\targ{}&\ctrl{1}&&\\
                \lstick{\ket{q_2}}& & & &\targ{}&&\\
                \lstick{\ket{coin}}&\gate{X}&\ctrl{-3}&\ctrl{-2}&\ctrl{-1}&\gate{X}&
            \end{quantikz}}
            \caption{The circuit on the left, is the increment circuit and the circuit on the right is the decrement circuit. After a coin toss, both circuits need to be employed. We need an increment for a rightwards shift and a decrement for a leftwards shift.}
            \label{fig:incdeccomponents}
        \end{figure}
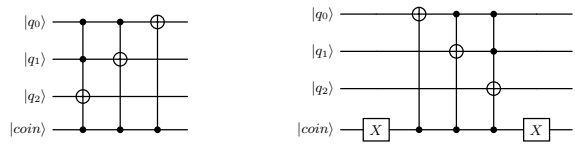
        \begin{figure}[!h]
            \centering
             \scalebox{.75}{
            \begin{quantikz}
                \lstick{\ket{q_0}}&&\ctrl{1}&\ctrl{1}&\targ{}& &\targ{}&\ctrl{1}&\ctrl{1}&&\\
                \lstick{\ket{q_1}}& &\ctrl{1}&\targ{}&& & &\targ{}&\ctrl{1}&&\\
                \lstick{\ket{q_2}}& &\targ{}& && & & &\targ{}&&\\
                \lstick{\ket{coin}}&\gate{H}&\ctrl{-1}&\ctrl{-2}&\ctrl{-3}&\gate{X}&\ctrl{-3}&\ctrl{-2}&\ctrl{-1}&\gate{X}&
            \end{quantikz}}
            \caption{Naive circuit for a single step of DTQW.}
            \label{fig:naivecircuit}
        \end{figure}
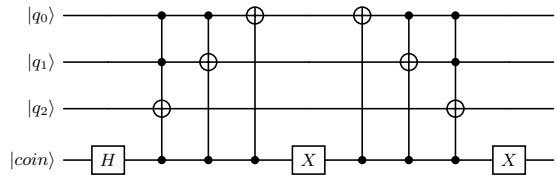
        \begin{table}[!h]
            \centering
            \begin{tabular}{|c|c|}
                \hline$\ket{x=0}=\ket{000}$ & \\\hline
                $\ket{x=-1}=\ket{111}$ & $\ket{x=+1}=\ket{001}$ \\\hline
                $\ket{x=-2}=\ket{110}$ & $\ket{x=+2}=\ket{010}$ \\\hline
                $\ket{x=-3}=\ket{101}$ & $\ket{x=+3}=\ket{011}$ \\\hline
            \end{tabular}
            \caption{Position state mapping for naive implementation of DTQW shown in Figure \ref{fig:naivecircuit}.}
            \label{tab:naive_mapping}
        \end{table}

\section{Our proposed approach: Enhanced increment-decrement approach for DTQW in binary quantum systems} \label{our-walk-approach}
    \subsection{Mathematical formulation of our proposed approach}
     As an instance, DTQW is applied over a 1D integer line. A coin is tossed and according to that, the walker shifts its position. Let $H_c$ be the Hilbert space of the coin spanned by the basis set $\{\ket{\uparrow},\ket{\downarrow}\}$ and $H_x$ be the Hilbert space of the position state by the basis set $\{\ket{x}\}_{x\in \mathbb{Z}}$. $\ket{x}$. $\ket{x}$ can be represented by multiple qubits like $\ket{q_n q_{n-1}...q_0 }$ where $q_n$ is the MSB and $q_0$ is the LSB.
        \newline
        In this approach two unitary operators are needed for the walk, unlike the naive approach, one operator is used for the even position state and another one is used for the odd position state. The operators are
        \begin{equation}
            U_{even}=S_{even}(C\otimes I)
        \end{equation}
        \begin{equation}
            U_{odd}=S_{odd}(C\otimes I)
        \end{equation}
        where $C$ is the coin toss operator and $S_{even}$ and $S_{odd}$ are the shift operators that are applied to the walker after the toss. A tensor product is denoted by $\otimes$.
        \newline
        The unitary operator for shifting the walker to an even position state is defined as
         \begin{equation}
            S_{even}=\sum_{x\in\mathbb{Z}}{(\ket{\uparrow}\bra{\uparrow}\otimes \ket{x_{e1}}\bra{x}+\ket{\downarrow}\bra{\downarrow}\otimes\ket{x_{e2}}\bra{x})}
        \label{shiftoperator_even}
        \end{equation}
        \newline
        where $\ket{x_{e1}}$ consists of $q_nq_{n-1}...q_0$. It can be written as $\ket{x_{n-1}+1}\ket{x_0+1}$ where $x_{n-1}$ consists of $q_nq_{n-1}...q_1$ and $x_0$ consists of $q_0$.
        \newline
        $\ket{x_{e2}}$ consists of $q_nq_{n-1}...q_0$. It can be written as $\ket{x_{n-1}}\ket{x_0+1}$ where $x_{n-1}$ consists of $q_nq_{n-1}...q_1$ and $x_0$ consists of $q_0$.
        \newline
        The unitary operator for shifting the walker to an odd position state is defined as
         \begin{equation}
            S_{odd}=\sum_{x\in\mathbb{Z}}{(\ket{\uparrow}\bra{\uparrow}\otimes \ket{x_{o1}}\bra{x}+\ket{\downarrow}\bra{\downarrow}\otimes\ket{x_{o2}}\bra{x})}
        \label{shiftoperator_odd}
        \end{equation}
        \newline
        where $\ket{x_{o1}}$ consists of $q_nq_{n-1}...q_0$. It can be written as $\ket{x_{n-1}}\ket{x_0-1}$ where $x_{n-1}$ consists of $q_nq_{n-1}...q_1$ and $x_0$ consists of $q_0$.
        \newline
        $\ket{x_{o2}}$ consists of $q_nq_{n-1}...q_0$. It can be written as $\ket{x_{n-1}-1}\ket{x_0-1}$ where $x_{n-1}$ consists of $q_nq_{n-1}...q_1$ and $x_0$ consists of $q_0$.
    
    \subsection{Implementation of the proposed approach}
       
\begin{figure}[!h]
                \centering
                 \scalebox{.5}{
                \begin{quantikz}
                    \lstick{\ket{q_0}} & &\gate{X}&& \\
                    \lstick{\ket{q_1}} & &\ctrl{1}&\targ{}& \\
                    \lstick{\ket{q_2}} & &\targ{}& & \\
                    \lstick{\ket{coin}} &\gate{H}&\ctrl{-1}&\ctrl{-2}& \\
                \end{quantikz}
                \hspace{2cm}
                \begin{quantikz}
                    \lstick{\ket{q_0}} & &\gate{X}&&&& \\
                    \lstick{\ket{q_1}} & & &\targ{}&\ctrl{1}&& \\
                    \lstick{\ket{q_2}} & & & &\targ{}&& \\
                    \lstick{\ket{coin}} &\gate{H} &\gate{X}&\ctrl{-2}&\ctrl{-1}&\gate{X}& \\
                \end{quantikz}}
                \caption{Circuit implementing DTQW in a 3-qubit system. The circuit on the left is for when the walker is in some even position state and the circuit on the right is for when the walker is in some odd position state.}
                \label{fig:our_technique_3qubits}
            \end{figure}
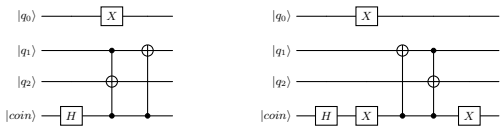

\begin{figure}[!h]
            \centering
            \includegraphics[width=.5\textwidth]{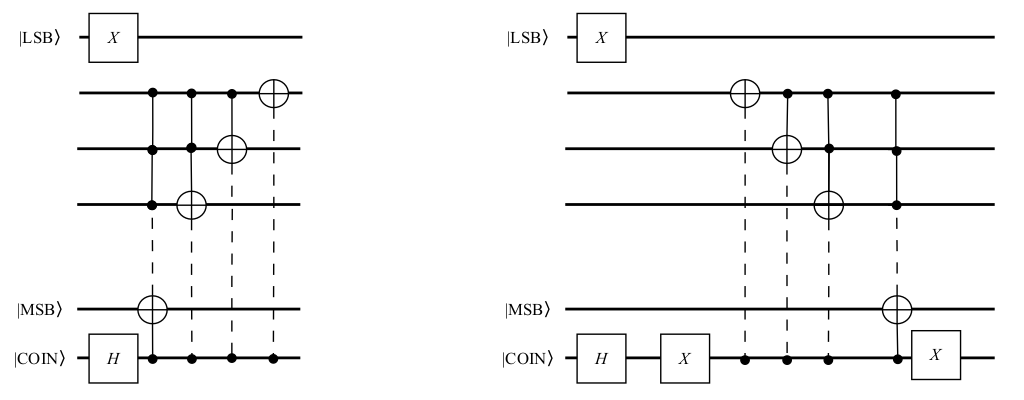}
            \caption{Generalized circuit diagram of our proposed enhanced increment decrement quantum walk for $n$ qubits. The circuit on the left is to be used when the walker is in some even position state and the circuit on the right is to be used in case the walker on some odd position state.}
            \label{fig:generalized-enh-inc-dec-ckt}
        \end{figure}

        In the naive implementation of the quantum walk, two components were required for increment and decrement in each coin toss. Inspired by the work in \cite{shivanisingh}, in our proposed algorithm, only one component is required in each coin toss. Figure \ref{fig:our_technique_3qubits} shows both components for a 3-qubit system. Extending the circuit for more than 3 qubits requires more multi-controlled Toffoli gates similar to the naive approach. Figure \ref{fig:generalized-enh-inc-dec-ckt} shows the generalized circuit diagram. It can be easily observed that, in our proposed algorithm, Algorithm \ref{algo:our_proposed_approach}, we can cut down the depth of the circuit for a single step of DTQW by half.

        \begin{algorithm}[H]
        \begin{algorithmic}[1]
                \caption{Our proposed approach of DTQW for binary systems}\label{algo:our_proposed_approach}
                \State $X \gets \text{$n+1$ qubit register}$ \Comment{Where the first $n$ qubits are to represent the walker's position state and the remaining one MSB is for the coin}
                \State $C \gets \text{A single qubit coin operator (Hadamard in our case)}$
                \State $S_{even} \gets \text{Shift operator as per Eq. \ref{shiftoperator_even}}$
                \State $S_{odd} \gets \text{Shift operator as per Eq. \ref{shiftoperator_odd}}$
                \State $N \gets 2^{n-1}-1$ \Comment{Total number of possible steps for a $n$-qubit quantum system}
                \State $i \gets 0$
                \While{$i < N$}
                    \State $X \gets (C \otimes I^{\otimes (n-1)}) \times X$ \Comment{Coin toss}
                    \If{$i$ is even} \Comment{Applying shift operation conditionally}
                        \State $X \gets (I \otimes S_{even}) \times X$
                    \Else
                        \State $X \gets (I \otimes S_{odd}) \times X$
                    \EndIf
                        
                    $i \gets i+1$
                \EndWhile
        \end{algorithmic}
        \end{algorithm} 
        \subsubsection{Explanation}
         In our approach, the basic concept of increment and decrement (shown in Figure \ref{fig:incdeccomponents}) is used with some modification. Indeed, it's widely understood that the increment and decrement circuit can be extended to $n$ qubits. In our proposed approach, each step incorporates either the enhanced increment or the enhanced decrement circuit (Shown in Figure \ref{fig:generalized-enh-inc-dec-ckt}).
         \begin{itemize}
             \item \textbf{The enhanced increment operation:} The walk starts at the initial state of $\ket{0}^{\otimes n}$. Following the coin toss, executed through a simple Hadamard transform, we proceed to select a single shift operator. Referring back to our discussion in Figure \ref{fig:generalized-enh-inc-dec-ckt}, since the walker currently resides in an even position state, our selection will be the enhanced increment circuit. This circuit performs an increment operation on every qubit except for the least significant bit (LSB) when the coin state happens to be \ket{1}. As an example, if we consider a 3 qubit system, then from $\ket{000}$ we get two states, $\ket{001}$ when the coin is $\ket{0}$, and $\ket{011}$ when the coin is $\ket{1}$. We can consider $\ket{011}$ as $+1$ state and $\ket{001}$ as $-1$ state. From this example, it's evident that for even position states, the least significant bit (LSB) will always be $\ket{0}$, while for odd position states, it will be $\ket{1}$.
             \item \textbf{The enhanced decrement operation:}
             After the first toss, the walker is on $\ket{001}$ and $\ket{011}$. Now we choose the enhanced decrement circuit as the shift operator. Akin to the enhanced increment circuit, it decrements all the qubits except for the LSB. So, from $\ket{001}$, we get $\ket{110}$ (a new state) when the coin is $\ket{0}$ and $\ket{000}$ (the previous state) when the coin is $\ket{1}$. We can consider the new state $\ket{110}$ as the $-2$ state. Similarly, from $\ket{011}$, we get $\ket{000}$ (the previous state) when the coin is $\ket{0}$ and $\ket{010}$ (a new state) when the coin is $\ket{1}$. We can consider the new state $\ket{010}$ as the $+2$ state.
         \end{itemize}

         The position state mapping employing the naive approach, shown in Table \ref{tab:naive_mapping}, uses binary 2’s complement counting scheme. Similarly, in our proposed approach, for all even position states, the LSB is always \ket{0}, and \ket{1} for all odd position states. This implies that the value of LSB always flips at every step of the DTQW. We have exploited this fact in our proposed approach and removed the flipping of the LSB in control with the coin state. In our approach, as it is also portrayed by the diagram (Fig. \ref{fig:generalized-enh-inc-dec-ckt}), the LSB is flipped irrespective of the coin state. For a 3-qubit system, the position state mapping employing our approach is shown in Table \ref{tab:our_technique_mapping_3qubits}.


          \begin{table}[!h]
                \centering
                \begin{tabular}{|c|c|}
                    \hline$\ket{x= 0}=\ket{000}$ & \\\hline
                    $\ket{x=-1}=\ket{001}$ & $\ket{x=+1}=\ket{011}$ \\\hline
                    $\ket{x=-2}=\ket{110}$ & $\ket{x=+2}=\ket{010}$ \\\hline
                    $\ket{x=-3}=\ket{111}$ & $\ket{x=+3}=\ket{101}$ \\\hline
                \end{tabular}
                \caption{Position state mapping employing our proposed approach for a 3-qubit system.}
                \label{tab:our_technique_mapping_3qubits}
            \end{table}

        \subsubsection{Mathematical steps for a 3-qubit DTQW using our proposed approach }We have taken \ket{000} as the initial position state of the walker. A total of three steps are as follows:
        \newline
        $a_0$\ket{0}\ket{000}.
        \newline
        $\downarrow U_{even}$
        \newline
        $a_1$\ket{0}\ket{001} + $b_1$\ket{1}\ket{011}.
        \newline
        $\downarrow U_{odd}$
        \newline
           $a_2$\ket{0}\ket{110} + $c_0$\ket{1}\ket{000} 
         + $c_1$\ket{0}\ket{000} + $b_2$\ket{1}\ket{010}.
         \newline
        $\downarrow U_{even}$
        \newline
           $a_3$\ket{0}\ket{111} + $c_{30}$\ket{1}\ket{001} 
         + $c_{10}$\ket{0}\ket{001} + $c_{11}$\ket{1}\ket{011} 
         + $c_{20}$\ket{0}\ket{001} + $c_{21}$\ket{1}\ket{011} 
         + $c_{31}$\ket{0}\ket{011} + $b_3$\ket{1}\ket{101}.
where $a$, $b$, and $c$ are the amplitudes.

 \subsection{Experimental results in binary systems}  \label{results}  
    \subsubsection{Comparative study between naive approach and our proposed approach} \label{naive-vs-ours}
 
   We consider three metrics for the comparison, which are as follows:     
        \begin{enumerate}
            \item \textbf{Circuit depth:} It is an important property of a quantum circuit. The circuit depth of a quantum circuit is the measure of how many layers of quantum gates, executed in parallel, it takes to complete the total computation. The depth of a circuit roughly corresponds to the amount of time it takes the quantum computer to execute the circuit. Therefore, we always want to reduce the depth of our quantum circuits as much as possible.
            \item \textbf{Gate count:} In Figure \ref{fig:naivecircuit} and Figure \ref{fig:generalized-enh-inc-dec-ckt}, we observe that multi-controlled Toffoli gates are used in abundance and for circuits with increasing number of qubits, this number of multi-controlled Toffoli's will increase substantially. However, these gates are not directly implementable on any quantum hardware. We need to decompose these gates into single and two-qubit gates before implementing them on real hardware. Gate count is the total number of gates after decomposing Toffoli gates. We also want to minimize the gate count of our circuits as much as possible. It is also possible to use intermediate qudits to decompose the Toffoli gates as suggested by the authors in \cite{sahaintermediate} for ancilla-free decomposition, which is thoroughly discussed in the next section.
            \item \textbf{Probability of success:} After the multi-controlled Toffoli gates are decomposed, certainly the depth and gate count of the circuit increase because we are replacing one gate with more than one other gate. We want a higher probability of success for our circuits. Small errors in the gates employed in quantum circuits are frequently present; these errors can be represented as an ideal gate followed by an unwanted Pauli operator. Instead of comparing the probability of small circuit errors, we compare the probability that the circuit will stay error-free (probability of success) in this comparison, maintaining the same level of generality as in \cite{majumdar2021optimizing}. The product that each component fails independently is the generalized formula for the probability of success ($P_{success}$) for any decomposition. The general equation for calculating success probability is,
            \begin{equation}\label{prob_success}
                P_{success}=\prod_{gates}{((P_{\text{success of gate}})^{\#gates}\times e^{-depth/T_1})}
            \end{equation}
            where $P_{success}$ is the success probability of the decomposed circuit, $P_{\text{success of gate}}$ is the success probability of each type of gate (single qubit/qudit or multi-qubit/qudit) as per the given hardware, $depth$ is the depth of the circuit after decomposition and $T_1$ is the relaxation time of quantum hardware. For further experiments, we take the value of $T_1$ as per \cite{https://doi.org/10.48550/arxiv.2203.07369}.
        \end{enumerate}

The comparative resource estimation for the naive approach and our proposed approach is exhibited in Table \ref{tab:generalized_resource_estimation_vs_naive}.

            \begin{table}[!h]
            \centering
            \scalebox{.55}{
            \begin{tabular}{|c|c|c|}
                   \hline \textbf{Topic} & \textbf{Our proposed approach} & \textbf{Naive approach \cite{douglas2009efficient}} \\
                   \hline 1-qubit Gate Count & $3t$ & $3t$ \\
                   \hline 2-qubit Gate Count & $t{(\sum^n_{x=3} (2n^2-6n+5)+1)}$ & $2t{(\sum^{n+1}_{x=3} (2n^2-6n+5)+2)}$\\
                   \hline Circuit Depth & $\frac{t}{2}{(\sum^n_{x=3} (8n-20)+1+1)}$ + $\frac{t}{2}{(\sum^n_{x=3} (8n-20)+1+3)}$ & $t{(2(\sum^{n+1}_3 (8n-20))+3)}$\\
                   \hline Highest Control & $(n-1)$-control & $n$-control\\
                   \hline
                \end{tabular}}
                \caption{Generalized formulation ($n$ - no. of qubit, $t$ - no. of steps) [Toffoli decomposition using 1-qubit and 2-qubit gates \cite{PhysRevA.87.062318}].}
                \label{tab:generalized_resource_estimation_vs_naive}
            \end{table}
 
    If we roughly estimate, the naive approach requires around $2n$ multi-controlled Toffoli gates for $n$ qubits, whereas our proposed approach requires roughly half that amount. Not only is the gate count approximately halved compared to the naive approach, but the circuit depth is also roughly halved.

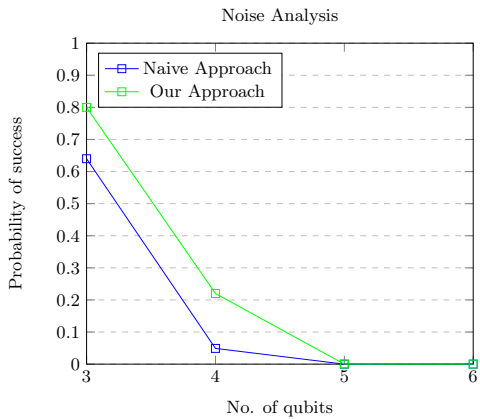
\begin{figure}[!h]
            \centering
            \scalebox{.75}{
                \begin{tikzpicture}
                    \begin{axis}[
                    title={Noise Analysis},
                        xlabel={No. of qubits},
                        ylabel={Probability of success},
                        xmin=3, xmax=6,
                        ymin=0, ymax=1,
                        xtick={3,4,5,6},
                        ytick={0,0.1,0.2,0.3,0.4,0.5,0.6,0.7,0.8,0.9,1},
                        legend pos=north west,
                        ymajorgrids=true,
                        grid style=dashed,
                    ]
    
                    \addplot[
                        color=blue,
                        mark=square,
                        ]
                        coordinates {
                        (3,0.64)(4,0.049)(5, 2.4642105893343263e-08)(6,1.2101035071989124e-41)
                        };
                    \addlegendentry{Naive Approach}
                    
                     \addplot[
                        color=green,
                        mark=square,
                        ]
                        coordinates {
                        (3,0.80)(4,0.22)(5,0.00015)(6,3.473613576228975e-21)
                        };
                    \addlegendentry{Our Approach}
                    \end{axis}
                \end{tikzpicture}}
                \caption{Success probability analysis between our proposed approach and the naive approach of DTQW.}
                \label{fig:success_prob_analysis_vs_naive}
            \end{figure}

Compared to the naive approach, our proposed approach provides a better probability of success. The result is as Figure \ref{fig:success_prob_analysis_vs_naive}.

    \subsubsection{Runs and Simulations}
        We use IBM's Qiskit package and the BasicAer backend to simulate and validate our approach. We decompose the circuit into 1-qubit and 2-qubit gates. We use Qiskit's 'transpile' function for decomposition. For decomposition, the basic gate set is ['cx', 'u1', 'u2', 'h', 'x', 'p']. In this set, only 'cx'(CNOT) is the 2-qubit gate, others are 1-qubit gates. The optimization level is set to 1. We also use ancilla qubits to decompose multi-qubit gates for better circuit depth. The total number of shots considered is 1024.
        
\begin{figure}
                \centering
                \scalebox{.75}{
                \begin{tikzpicture}
                    \begin{axis}[
                        xlabel=States,
                        xtick=data,
                        ylabel=Probabilities,,
                	ybar,
                    ]
                        \addplot coordinates{(-3, 0.1220703125) (1, 0.6328125) (3, 0.1103515625) (-1, 0.134765625)};
                    \end{axis}

                \end{tikzpicture}}
                \caption{Probability distribution of the simulated result for 3-qubit DTQW using our proposed approach.}
                \label{fig:our_technique_3qubits_histogram}
            \end{figure}
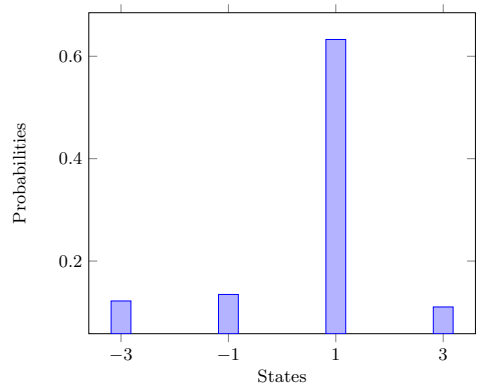

        The results of DTQW using our proposed approach are as follows:
        
        \paragraph{3-qubit quantum walks using our proposed approach}
            We have implemented a 3-qubit quantum walk using our proposed approach. The circuits are shown in Figure \ref{fig:our_technique_3qubits} and the corresponding position state mapping is as Table \ref{tab:our_technique_mapping_3qubits}. The probability distribution after three steps of quantum walks is portrayed in Figure \ref{fig:our_technique_3qubits_histogram}.

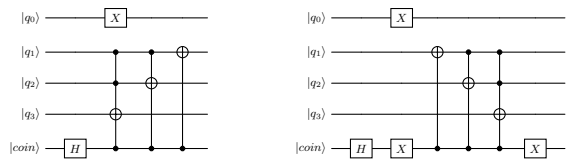
\begin{figure}[!h]
            \centering
             \scalebox{.5}{
            \begin{quantikz}
                \lstick{\ket{q_0}} & &\gate{X}&&& \\
                \lstick{\ket{q_1}} & &\ctrl{1}&\ctrl{1}&\targ{}& \\
                \lstick{\ket{q_2}} & &\ctrl{1}&\targ{}& & \\
                \lstick{\ket{q_3}} & &\targ{} & & & \\
                \lstick{\ket{coin}} &\gate{H}&\ctrl{-1}&\ctrl{-2}&\ctrl{-3}& \\
            \end{quantikz}

        \hspace{2cm}
            \begin{quantikz}
                \lstick{\ket{q_0}} & &\gate{X}&&&&& \\
                \lstick{\ket{q_1}} & & &\targ{}&\ctrl{1}&\ctrl{1}&& \\
                \lstick{\ket{q_2}} & & & &\targ{}&\ctrl{1}&& \\
                \lstick{\ket{q_3}} & & & & &\targ{}&& \\
                \lstick{\ket{coin}} &\gate{H}&\gate{X}&\ctrl{-3}&\ctrl{-2}&\ctrl{-1}&\gate{X}& \\
            \end{quantikz}}
            \caption{Circuit implementing DTQW in a 4-qubit system. The circuit on the left is for when the walker is in some even position state and the circuit on the right is for when the walker is in some odd position state.}
            \label{fig:our_technique_4qubits}
            \end{figure}

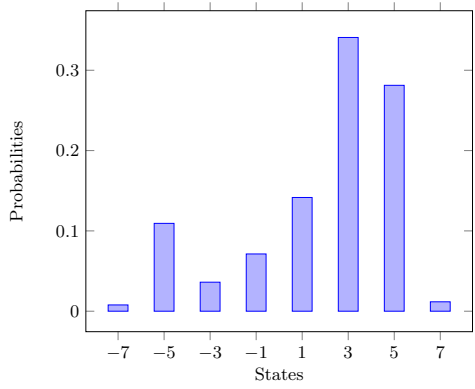
\begin{figure}
                \centering
                \scalebox{.75}{
                \begin{tikzpicture}
                    \begin{axis}[
                        xlabel=States,
                        xtick=data,
                        ylabel=Probabilities,,
                	ybar,
                    ]
                        \addplot coordinates{(-3, 0.0361328125) (-7, 0.0078125) (7, 0.01171875) (-5, 0.109375) (-1, 0.0712890625) (5, 0.28125) (1, 0.1416015625) (3, 0.3408203125)};
                    \end{axis}

                \end{tikzpicture}}
                \caption{Probability distribution of the simulated result for 4-qubit DTQW using our proposed approach.}
                \label{fig:our_technique_4qubits_histogram}
            \end{figure}

        \paragraph{4-qubit Quantum Walk Using Our Proposed approach}
            For a 4-qubit system, the circuits of the DTQW are shown in Figure \ref{fig:our_technique_4qubits} and the corresponding position state mapping is as Table \ref{tab:our_technique_mapping_4qubits}. The probability distribution after seven steps of quantum walks is portrayed in Figure \ref{fig:our_technique_4qubits_histogram}.

            \begin{table}[!h]
                \centering
                \begin{tabular}{|c|c|}
                    \hline$\ket{x= 0}=\ket{0000}$ & \\\hline
                    $\ket{x=-1}=\ket{0001}$ & $\ket{x=+1}=\ket{0011}$ \\\hline
                    $\ket{x=-2}=\ket{1110}$ & $\ket{x=+2}=\ket{0010}$ \\\hline
                    $\ket{x=-3}=\ket{1111}$ & $\ket{x=+3}=\ket{0101}$ \\\hline
                    $\ket{x=-4}=\ket{1100}$ & $\ket{x=+4}=\ket{0100}$ \\\hline
                    $\ket{x=-5}=\ket{1101}$ & $\ket{x=+5}=\ket{0111}$ \\\hline
                    $\ket{x=-6}=\ket{1010}$ & $\ket{x=+6}=\ket{0110}$ \\\hline
                    $\ket{x=-7}=\ket{1011}$ & $\ket{x=+7}=\ket{1001}$ \\\hline
                \end{tabular}
                \caption{Position state mapping employing our proposed approach for a 4-qubit system.}
                \label{tab:our_technique_mapping_4qubits}
            \end{table}
    

   These results validate our proposed approach and exhibit efficiency as compared to the naive approach.

    \section{enhanced increment-decrement approach for DTQW in higher dimensional quantum systems} \label{our-walk-approach-higher-dimension}
        In this section, at first, we try to map our proposed binary method to higher dimensional quantum systems by considering LSB is in the binary system. Therefore, for $n$ qudits with dimension $d$, we can have a total of $2d^{(n-1)}$ number of state spaces and we can implement a total of $d^{n-1}$ number of steps. Thus implementing our proposed methodology in higher dimensions is limiting us from leveraging the whole state space of higher dimensions because of the LSB which is set to $d=2$.

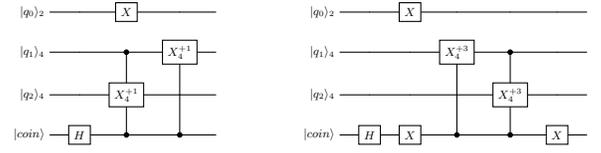
\begin{figure} [!h]
            \centering
            \scalebox{.5}{
            \begin{quantikz}
                \lstick{\ket{q_0}$_2$} & &\gate{X}&& \\
                \lstick{\ket{q_1}$_4$} & &\ctrl{1}&\gate{X^{+1}_{4}}& \\
                \lstick{\ket{q_2}$_4$} & &\gate{X^{+1}_{4}}& & \\
                \lstick{\ket{coin}} &\gate{H}&\ctrl{-1}&\ctrl{-2}& \\
            \end{quantikz}
            \hspace{2cm}
            \begin{quantikz}
                \lstick{\ket{q_0}$_2$} & &\gate{X}&&&& \\
                \lstick{\ket{q_1}$_4$} & & &\gate{X^{+3}_{4}}&\ctrl{1}&& \\
                \lstick{\ket{q_2}$_4$} & & & &\gate{X^{+3}_{4}}&& \\
                \lstick{\ket{coin}} &\gate{H} &\gate{X}&\ctrl{-2}&\ctrl{-1}&\gate{X}& \\
            \end{quantikz}}
            \caption{The enhanced increment decrement approach for a 3-ququad system.}
            \label{fig:our_technique_higher_dimension_3ququads}
        \end{figure}

            As an instance, in a ququad system (Four-dimensional quantum systems), the circuit for the 3-ququad DTQW is as Figure \ref{fig:our_technique_higher_dimension_3ququads}, and the corresponding position state mapping is shown in Table \ref{tab:our_technique_mapping_higher_dimension_3ququads}. In the last step of Table \ref{tab:our_technique_mapping_higher_dimension_3ququads}, we get the states \ket{200} as the mid-state of the total state set in both directions,  which restricts the walker to use the rest of the 32 number of state spaces further.  We also check upto 8-dimensional quantum systems and observe that this direct mapping technique in higher dimensions can not be generalized for all $n$-qudit systems since it can not use all possible state spaces. Therefore, we need some changes in our proposed technique of binary systems while applying in higher dimensions so that it also utilizes all the state spaces in the higher dimensional quantum systems.
    
        \begin{table}[!h]
            \centering
            \begin{tabular}{|c|c|}
                \hline$\ket{x= 0}=\ket{000}$ &  \\\hline
                $\ket{x=-1}=\ket{001}$ & $\ket{x=+1}=\ket{011}$ \\\hline
                $\ket{x=-2}=\ket{330}$ & $\ket{x=+2}=\ket{010}$ \\\hline
                $\ket{x=-3}=\ket{331}$ & $\ket{x=+3}=\ket{021}$ \\\hline
                $\ket{x=-4}=\ket{320}$ & $\ket{x=+4}=\ket{020}$ \\\hline
                $\ket{x=-5}=\ket{321}$ & $\ket{x=+5}=\ket{031}$ \\\hline
                $\ket{x=-6}=\ket{310}$ & $\ket{x=+6}=\ket{030}$ \\\hline
                $\ket{x=-7}=\ket{331}$ & $\ket{x=+7}=\ket{101}$ \\\hline
                $\ket{x=-8}=\ket{300}$ & $\ket{x=+8}=\ket{100}$ \\\hline
                $\ket{x=-9}=\ket{301}$ & $\ket{x=+9}=\ket{111}$ \\\hline
                $\ket{x=-10}=\ket{230}$ & $\ket{x=+10}=\ket{110}$ \\\hline
                $\ket{x=-11}=\ket{231}$ & $\ket{x=+11}=\ket{121}$ \\\hline
                $\ket{x=-12}=\ket{220}$ & $\ket{x=+12}=\ket{120}$ \\\hline
                $\ket{x=-13}=\ket{221}$ & $\ket{x=+13}=\ket{131}$ \\\hline
                $\ket{x=-14}=\ket{210}$ & $\ket{x=+14}=\ket{130}$ \\\hline
                $\ket{x=-15}=\ket{211}$ & $\ket{x=+15}=\ket{201}$ \\\hline
                $\ket{x=-16}=\ket{200}$ & $\ket{x=+16}=\ket{200}$ \\\hline
            \end{tabular}
            \caption{Position state mapping of the enhanced increment decrement approach in a 3-ququad system shown in Figure \ref{fig:our_technique_higher_dimension_3ququads}.}
            \label{tab:our_technique_mapping_higher_dimension_3ququads}
        \end{table}

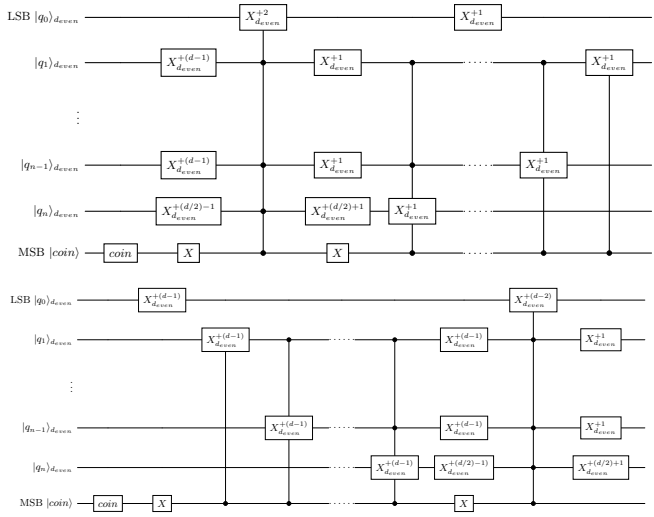
\begin{figure}[!h]
            \centering
            \scalebox{.5}{
            \begin{quantikz}
                \lstick{LSB \ket{q_0}$_{d_{even}}$} && &\gate{X^{+2}_{d_{even}}}&&&\gate{X^{+1}_{d_{even}}}&&& \\
                \lstick{\ket{q_1}$_{d_{even}}$} &&\gate{X^{+(d-1)}_{d_{even}}}&\ctrl{-1}&\gate{X^{+1}_{d_{even}}}&\ctrl{2}&\ldots\ldots&\ctrl{2}&\gate{X^{+1}_{d_{even}}}& \\[0.5cm]
                \lstick{\vdots} \\[0.5cm]
                \lstick{\ket{q_{n-1}}$_{d_{even}}$} & &\gate{X^{+(d-1)}_{d_{even}}}&\ctrl{0}&\gate{X^{+1}_{d_{even}}}&\ctrl{1}&\ldots\ldots&\gate{X^{+1}_{d_{even}}}&& \\
                \lstick{\ket{q_{n}}$_{d_{even}}$} & &\gate{X^{+(d/2)-1}_{d_{even}}}&\ctrl{0}&\gate{X^{+(d/2)+1}_{d_{even}}}&\gate{X^{+1}_{d_{even}}}&\ldots\ldots& && \\
                \lstick{MSB \ket{coin}} &\gate{coin}&\gate{X}&\ctrl{-4}&\gate{X}&\ctrl{-1}&\ldots\ldots&\ctrl{-2}&\ctrl{-4}& \\
            \end{quantikz}}
            \scalebox{.43}{
            \begin{quantikz}
                \lstick{LSB \ket{q_0}$_{d_{even}}$} &&\gate{X^{+(d-1)}_{d_{even}}}&&&&&&\gate{X^{+(d-2)}_{d_{even}}}&& \\
                \lstick{\ket{q_1}$_{d_{even}}$} && &\gate{X^{+(d-1)}_{d_{even}}}&\ctrl{2}&\ldots\ldots&\ctrl{2}&\gate{X^{+(d-1)}_{d_{even}}}&\ctrl{-1}&\gate{X^{+1}_{d_{even}}}& \\[0.5cm]
                \lstick{\vdots} \\[0.5cm]
                \lstick{\ket{q_{n-1}}$_{d_{even}}$} && &&\gate{X^{+(d-1)}_{d_{even}}}&\ldots\ldots&\ctrl{1}&\gate{X^{+(d-1)}_{d_{even}}}&\ctrl{0}&\gate{X^{+1}_{d_{even}}}& \\
                \lstick{\ket{q_n}$_{d_{even}}$} && && &\ldots\ldots&\gate{X^{+(d-1)}_{d_{even}}}&\gate{X^{+(d/2)-1)}_{d_{even}}}&\ctrl{0}&\gate{X^{+(d/2)+1}_{d_{even}}}& \\
                \lstick{MSB \ket{coin}} &\gate{coin}&\gate{X}&\ctrl{-4}&\ctrl{-2}&\ldots\ldots&\ctrl{-1}&\gate{X}&\ctrl{-4}&& \\
            \end{quantikz}}
            \caption{Generalized modified DTQW circuit for all even dimensions of qudits. After the binary generic coin toss, the circuit on the top is to be applied when the walker is in an even position state. Otherwise, when the walker is in an odd position state, the circuit below is to be applied.}
            \label{fig:our_technique_higher_dimension_improved_generalized}
        \end{figure}

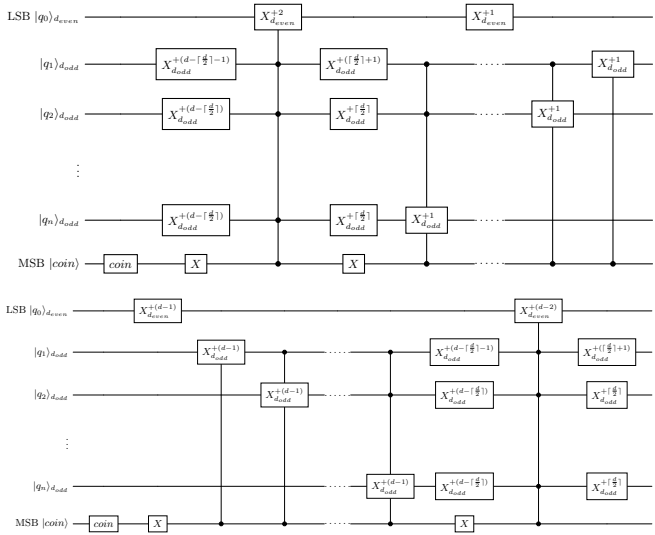
\begin{figure}[!h]
            \centering
            \scalebox{.5}{
            \begin{quantikz}
                \lstick{LSB \ket{q_0}$_{d_{even}}$} &&&\gate{X^{+2}_{d_{even}}}&&&\gate{X^{+1}_{d_{even}}}&&& \\
                \lstick{\ket{q_1}$_{d_{odd}}$} &&\gate{X^{+(d-\lceil\frac{d}{2}\rceil-1)}_{d_{odd}}}&\ctrl{0}&\gate{X^{+(\lceil\frac{d}{2}\rceil+1)}_{d_{odd}}}&\ctrl{1}&\ldots\ldots&\ctrl{1}&\gate{X^{+1}_{d_{odd}}}& \\
                \lstick{\ket{q_{2}}$_{d_{odd}}$} &&\gate{X^{+(d-\lceil\frac{d}{2}\rceil)}_{d_{odd}}}&\ctrl{0}&\gate{X^{+\lceil\frac{d}{2}\rceil}_{d_{odd}}}&\ctrl{2}&\ldots\ldots&\gate{X^{+1}_{d_{odd}}}&& \\[0.5cm]
                \lstick{\vdots} \\[0.5cm]
                \lstick{\ket{q_n}$_{d_{odd}}$} &&\gate{X^{+(d-\lceil\frac{d}{2}\rceil)}_{d_{odd}}}&\ctrl{0}&\gate{X^{+\lceil\frac{d}{2}\rceil}_{d_{odd}}}&\gate{X^{+1}_{d_{odd}}}&\ldots\ldots& && \\
                \lstick{MSB \ket{coin}} &\gate{coin}&\gate{X}&\ctrl{-5}&\gate{X}&\ctrl{-1}&\ldots\ldots&\ctrl{-3}&\ctrl{-4}& \\
            \end{quantikz}}
            \scalebox{.43}{
            \begin{quantikz}
                \lstick{LSB \ket{q_0}$_{d_{even}}$} &&\gate{X^{+(d-1)}_{d_{even}}}&&&&&&\gate{X^{+(d-2)}_{d_{even}}}&& \\
                \lstick{\ket{q_1}$_{d_{odd}}$} && &\gate{X^{+(d-1)}_{d_{odd}}}&\ctrl{1}&\ldots\ldots&\ctrl{1}&\gate{X^{+(d-\lceil\frac{d}{2}\rceil-1)}_{d_{odd}}}&\ctrl{-1}&\gate{X^{+(\lceil\frac{d}{2}\rceil+1)}_{d_{odd}}}& \\
                \lstick{\ket{q_2}$_{d_{odd}}$} && &&\gate{X^{+(d-1)}_{d_{odd}}}&\ldots\ldots&\ctrl{2}&\gate{X^{+(d-\lceil\frac{d}{2}\rceil)}_{d_{odd}}}&\ctrl{-1}&\gate{X^{+\lceil\frac{d}{2}\rceil}_{d_{odd}}}& \\[0.5cm]
                \lstick{\vdots} \\[0.5cm]
                \lstick{\ket{q_n}$_{d_{odd}}$} && && &\ldots\ldots&\gate{X^{+(d-1)}_{d_{odd}}}&\gate{X^{+(d-\lceil\frac{d}{2}\rceil)}_{d_{odd}}}&\ctrl{0}&\gate{X^{+\lceil\frac{d}{2}\rceil}_{d_{odd}}}& \\
                \lstick{MSB \ket{coin}} &\gate{coin}&\gate{X}&\ctrl{-4}&\ctrl{-3}&\ldots\ldots&\ctrl{-1}&\gate{X}&\ctrl{-3}&& \\
            \end{quantikz}}
            \caption{Generalized modified DTQW circuit for all odd dimensions of qudits. After the binary generic coin toss, the circuit on the top is to be applied when the walker is in an even position state. Otherwise, when the walker is in an odd position state, the circuit below is to be applied.}
            \label{fig:our_technique_higher_dimension_improved_generalized_odd}
        \end{figure}

            To solve the state-space issue in higher dimensions, with a slight modification of the proposed binary-only approach, we propose a modified technique where we set the Least Significant Bit (LSB) as any even dimension when $d \ge 4$. We propose modified circuits for both the even and odd dimensions when $d > 2$, which are portrayed in Figures \ref{fig:our_technique_higher_dimension_improved_generalized} and \ref{fig:our_technique_higher_dimension_improved_generalized_odd} respectively. 
            

           In our modified circuit, as shown in Figures \ref{fig:our_technique_higher_dimension_improved_generalized} and \ref{fig:our_technique_higher_dimension_improved_generalized_odd}, instead of applying only a NOT/Pauli-X gate on the LSB, a conditional increment or decrement gate is applied on the LSB for both the increment and decrement circuits so that the same mid-state of the total state set can be avoided in both directions of DTQW. With the modified logic as per Algorithm \ref{algo:higherdimension}, when the walker reaches the middle point, it increments the value of LSB by 2 because of the increment circuit and decrements the value of LSB by 2 because of the decrement circuit. Hence, with this modified circuit, all possible states can be traversed using the even-dimensional circuit, which is $d^{n}$ and the total number of steps that can be implemented is $(d^{n}/2)-1$. In our modified circuit, the LSB is not only dependent on control operations from other qudits for a conditional increment or decrement operation, but also it must increment the value of LSB at one step, and in the very next step, it decrements the value of LSB. For that reason, we need to restrict the LSB as an even dimension, but the other qudits can be in any dimension. Henceforth, the total number of states that can be traversed using the odd-dimensional circuit is $d_{even}*d^{n-1}$ and the total number of steps that can be implemented is $((d_{even}*d^{n-1})/2)-1$, where $d_{even}$ represents the LSB.

         \begin{algorithm}[H]
        \caption{Implementing DTQW in higher dimensional ($>2$) quantum systems using our proposed approach}
        \begin{algorithmic}
            \State $X \gets \text{Position State ($n$-qudit of $d$ dimension)}$
            \State $S_{even} \gets \text{Shift Operator as per Eq. \ref{shiftoperator_even}}$
            \State $S_{odd} \gets \text{Shift Operator as per Eq. \ref{shiftoperator_odd}}$
            \State $C \gets \text{Coin Operator ( 1-qubit Hadamard Coin)}$
            \State $steps \gets \text{Half of the total number of possible states for X}$
            \State $step \gets 0$
            \While{$step \leq $steps}
                \State $\text{Apply C over X i.e., } C \otimes X$
                \If{$step$ is even}
                    \State $\text{Apply }S_{even}\text{ over }C \otimes X $ \Comment{C controls Shift Operator and it works over X}
                \ElsIf{$step$ is odd}
                    \State $\text{Apply }S_{odd}\text{ over }C \otimes X $ \Comment{C controls Shift Operator and it works over X}
                \EndIf
                \If{$X = mid point (excluding\_LSB)$ and $step$ is odd}
                    \State $\text{Apply conditional decrement operator.}$ \Comment{Fig
                    [\ref{fig:our_technique_higher_dimension_improved_generalized}] or [\ref{fig:our_technique_higher_dimension_improved_generalized_odd}] according to d}
                \ElsIf{$X = mid point (excluding\_LSB)$ and $step$ is even}
                    \State $\text{Apply conditional increment operator.}$ \Comment{Fig [\ref{fig:our_technique_higher_dimension_improved_generalized}] or [\ref{fig:our_technique_higher_dimension_improved_generalized_odd}] according to d}
                \EndIf
                \State $step \gets step + 1$
            \EndWhile
        \end{algorithmic}
        \label{algo:higherdimension}
        \end{algorithm}  
           
           For better understanding, we take two separate example circuits, (i) an even dimension (3-ququad DTQW) circuit and (ii) an odd dimension (3-qutrit DTQW) circuit.

\begin{figure}[!h]
            \centering
            \scalebox{.65}{
            \begin{quantikz}
                \lstick{\ket{q_0}$_4$} &&&\gate{X^{+2}_4}&&\gate{X^{+1}_4}&& \\
                \lstick{\ket{q_1}$_4$} &&\gate{X^{+3}_4}&\ctrl{0}&\gate{X^{+1}_4}&\ctrl{1}&\gate{X^{+1}_4}& \\
                \lstick{\ket{q_2}$_4$} &&\gate{X^{+1}_4}&\ctrl{0}&\gate{X^{+3}_4}&\gate{X^{+1}_4}& & \\
                \lstick{\ket{coin}} &\gate{H} &\gate{X}&\ctrl{-3}&\gate{X}&\ctrl{-1}&\ctrl{-2}& \\
            \end{quantikz}}
            \scalebox{.65}{
            \begin{quantikz}
                \lstick{\ket{q_0}$_4$} & &\gate{X^{+3}_4}&&&&\gate{X^{+2}_4}&& \\
                \lstick{\ket{q_1}$_4$} & & &\gate{X^{+3}_4}&\ctrl{1}&\gate{X^{+3}_4}&\ctrl{0}&\gate{X^{+1}_4}& \\
                \lstick{\ket{q_2}$_4$} & & & &\gate{X^{+3}_4}&\gate{X^{+1}_4}&\ctrl{0}&\gate{X^{+3}_4}& \\
                \lstick{\ket{coin}} &\gate{H} &\gate{X}&\ctrl{-2}&\ctrl{-1}&\gate{X}&\ctrl{-3}&& \\
            \end{quantikz}}
            \caption{Implementation of the modified enhanced increment decrement approach for DTQW in a 3-ququad system.}
            \label{fig:our_technique_4_dimension_improved_3qudits}
        \end{figure}
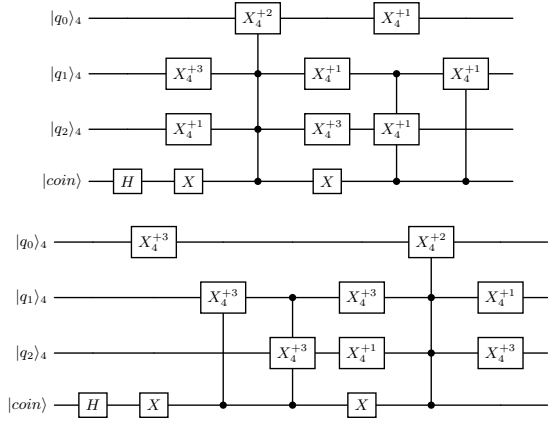

        
   \paragraph{Implementation of 3-ququad (even dimension) DTQW with a Hadamard coin:}
   As an example, we again take ququad systems, but this time with 3 ququads, including the LSB as ququad as per the modified circuit. Earlier, we observed in Table \ref{tab:our_technique_mapping_higher_dimension_3ququads}, at step 16, we get $\ket{200}$ in both the directions of DTQW. As per the circuit in Figure \ref{fig:our_technique_higher_dimension_3ququads}, we get these states because we have used only binary ($\ket{0}$ and $\ket{1}$) for the LSB. Since the modified enhanced increment-decrement approach for DTQW can utilize $\ket{2}$ and $\ket{3}$ states in the LSB as shown in Figure \ref{fig:our_technique_4_dimension_improved_3qudits}, we get the following states in Table \ref{tab:our_technique_mapping_4_dimension_improved_3qudits}. In Table \ref{tab:our_technique_mapping_4_dimension_improved_3qudits}, We observe that at step 16, $\ket{x=+16}=\ket{200}$ becomes $\ket{x=+16}=\ket{202}$ by incrementing the value of LSB by 2 with the help of controlled increment operator as per modified circuit's logic. After this step, the previous process of DTQW again starts working and we get all possible state spaces as required.

        \begin{table}[!h]
            \centering
            \begin{tabular}{|c|c|}
                \hline$\ket{x= 0}=\ket{000}$ &  \\\hline
                $\ket{x=-1}=\ket{001}$ & $\ket{x=+1}=\ket{011}$ \\\hline
                $\ket{x=-2}=\ket{330}$ & $\ket{x=+2}=\ket{010}$ \\\hline
                $\ket{x=-3}=\ket{331}$ & $\ket{x=+3}=\ket{021}$ \\\hline
                $\ket{x=-4}=\ket{320}$ & $\ket{x=+4}=\ket{020}$ \\\hline
                $\ket{x=-5}=\ket{321}$ & $\ket{x=+5}=\ket{031}$ \\\hline
                $\ket{x=-6}=\ket{310}$ & $\ket{x=+6}=\ket{030}$ \\\hline
                $\ket{x=-7}=\ket{331}$ & $\ket{x=+7}=\ket{101}$ \\\hline
                $\ket{x=-8}=\ket{300}$ & $\ket{x=+8}=\ket{100}$ \\\hline
                $\ket{x=-9}=\ket{301}$ & $\ket{x=+9}=\ket{111}$ \\\hline
                $\ket{x=-10}=\ket{230}$ & $\ket{x=+10}=\ket{110}$ \\\hline
                $\ket{x=-11}=\ket{231}$ & $\ket{x=+11}=\ket{121}$ \\\hline
                $\ket{x=-12}=\ket{220}$ & $\ket{x=+12}=\ket{120}$ \\\hline
                $\ket{x=-13}=\ket{221}$ & $\ket{x=+13}=\ket{131}$ \\\hline
                $\ket{x=-14}=\ket{210}$ & $\ket{x=+14}=\ket{130}$ \\\hline
                $\ket{x=-15}=\ket{211}$ & $\ket{x=+15}=\ket{201}$ \\\hline
                $\ket{x=-16}=\ket{200}$ & $\ket{x=+16}=\ket{202}$ \\\hline
                $\ket{x=-17}=\ket{203}$ & $\ket{x=+17}=\ket{213}$ \\\hline
                $\ket{x=-18}=\ket{132}$ & $\ket{x=+18}=\ket{212}$ \\\hline
                $\ket{x=-19}=\ket{133}$ & $\ket{x=+19}=\ket{223}$ \\\hline
                $\ket{x=-20}=\ket{122}$ & $\ket{x=+20}=\ket{222}$ \\\hline
                $\ket{x=-21}=\ket{123}$ & $\ket{x=+21}=\ket{233}$ \\\hline
                $\ket{x=-22}=\ket{112}$ & $\ket{x=+22}=\ket{232}$ \\\hline
                $\ket{x=-23}=\ket{113}$ & $\ket{x=+23}=\ket{303}$ \\\hline
                $\ket{x=-24}=\ket{102}$ & $\ket{x=+24}=\ket{302}$ \\\hline
                $\ket{x=-25}=\ket{103}$ & $\ket{x=+25}=\ket{313}$ \\\hline
                $\ket{x=-26}=\ket{032}$ & $\ket{x=+26}=\ket{312}$ \\\hline
                $\ket{x=-27}=\ket{033}$ & $\ket{x=+27}=\ket{323}$ \\\hline
                $\ket{x=-28}=\ket{022}$ & $\ket{x=+28}=\ket{322}$ \\\hline
                $\ket{x=-29}=\ket{023}$ & $\ket{x=+29}=\ket{333}$ \\\hline
                $\ket{x=-30}=\ket{012}$ & $\ket{x=+30}=\ket{332}$ \\\hline
                $\ket{x=-31}=\ket{013}$ & $\ket{x=+31}=\ket{003}$ \\\hline
                $\ket{x=-32}=\ket{002}$ & $\ket{x=+32}=\ket{002}$ \\\hline
            \end{tabular}
            \caption{Position state mapping of modified DTQW in a 3-ququad system shown in Figure \ref{fig:our_technique_4_dimension_improved_3qudits}.}
            \label{tab:our_technique_mapping_4_dimension_improved_3qudits}
        \end{table}

    \paragraph{Implementation of 3-qutrit (odd dimension) DTQW with a Hadamard coin:}

        We have also shown an example of our modified enhanced increment decrement approach for DTQW in a 3-qutrit system (Figure \ref{fig:our_technique_3_dimension_improved_3qudits}) and we get the results shown in Table \ref{tab:our_technique_mapping_odd_higher_dimension_improved_3qudits} by considering LSB as in ququad system. If we had used only our enhanced increment decrement circuit, we could get up to step 9, where we have to stop the DTQW with \ket{x=+9}=\ket{121} and \ket{x=-9}=\ket{121}. Because of our modified circuit, it can be well-avoided. The \ket{x=-9}=\ket{121} now becomes \ket{123} since the conditional increment activates at the midpoint as per the logic to get access to the full state spaces. 
    
        \begin{figure}[!h]
            \centering
            \scalebox{0.65}{
            \begin{quantikz}
                \lstick{\ket{q_0}$_4$} &&&\gate{X^{+2}_4}&&\gate{X^{+1}_4}&& \\
                \lstick{\ket{q_1}$_3$} &&&\ctrl{0}&&\ctrl{1}&\gate{X^{+1}_3}& \\
                \lstick{\ket{q_2}$_3$} &&\gate{X^{+1}_3}&\ctrl{0}&\gate{X^{+2}_3}&\gate{X^{+1}_3}& & \\
                \lstick{\ket{coin}} &\gate{H}&\gate{X}&\ctrl{-3}&\gate{X}&\ctrl{-1}&\ctrl{-2}& \\
            \end{quantikz}}
           \scalebox{.65}{
            \begin{quantikz}
                \lstick{\ket{q_0}$_4$} &&\gate{X^{+3}_4}&&&&\gate{X^{+2}_4}&& \\
                \lstick{\ket{q_1}$_3$} && &\gate{X^{+2}_3}&\ctrl{1}&&\ctrl{0}&& \\
                \lstick{\ket{q_2}$_3$} && & &\gate{X^{+2}_3}&\gate{X^{+1}_3}&\ctrl{0}&\gate{X^{+2}_3}& \\
                \lstick{\ket{coin}} &\gate{H}&\gate{X}&\ctrl{-2}&\ctrl{-1}&\gate{X}&\ctrl{-3}&& \\
            \end{quantikz}}
            \caption{Implementation of the modified enhanced increment decrement approach for DTQW in a 3-qutrit system.}
            \label{fig:our_technique_3_dimension_improved_3qudits}
        \end{figure}
            
        \begin{table}[!h]
            \centering
            \begin{tabular}{|c|c|}
                \hline$\ket{x= 0}=\ket{000}$ &  \\\hline
                $\ket{x=-1}=\ket{001}$ & $\ket{x=+1}=\ket{011}$ \\\hline
                $\ket{x=-2}=\ket{220}$ & $\ket{x=+2}=\ket{010}$ \\\hline
                $\ket{x=-3}=\ket{221}$ & $\ket{x=+3}=\ket{021}$ \\\hline
                $\ket{x=-4}=\ket{210}$ & $\ket{x=+4}=\ket{020}$ \\\hline
                $\ket{x=-5}=\ket{211}$ & $\ket{x=+5}=\ket{101}$ \\\hline
                $\ket{x=-6}=\ket{200}$ & $\ket{x=+6}=\ket{100}$ \\\hline
                $\ket{x=-7}=\ket{201}$ & $\ket{x=+7}=\ket{111}$ \\\hline
                $\ket{x=-8}=\ket{120}$ & $\ket{x=+8}=\ket{110}$ \\\hline
                $\ket{x=-9}=\ket{123}$ & $\ket{x=+9}=\ket{121}$ \\\hline
                $\ket{x=-10}=\ket{112}$ & $\ket{x=+10}=\ket{122}$ \\\hline
                $\ket{x=-11}=\ket{113}$ & $\ket{x=+11}=\ket{203}$ \\\hline
                $\ket{x=-12}=\ket{102}$ & $\ket{x=+12}=\ket{202}$ \\\hline
                $\ket{x=-13}=\ket{103}$ & $\ket{x=+13}=\ket{213}$ \\\hline
                $\ket{x=-14}=\ket{022}$ & $\ket{x=+14}=\ket{212}$ \\\hline
                $\ket{x=-15}=\ket{023}$ & $\ket{x=+15}=\ket{223}$ \\\hline
                $\ket{x=-16}=\ket{012}$ & $\ket{x=+16}=\ket{222}$ \\\hline
                $\ket{x=-17}=\ket{013}$ & $\ket{x=+17}=\ket{003}$ \\\hline
                $\ket{x=-18}=\ket{002}$ & $\ket{x=+18}=\ket{002}$ \\\hline
            \end{tabular}
            \caption{Position state mapping of modified DTQW in a 3-qutrit system shown in Figure \ref{fig:our_technique_3_dimension_improved_3qudits}.}
            \label{tab:our_technique_mapping_odd_higher_dimension_improved_3qudits}
        \end{table}

\subsection{Decomposition of Multi-controlled Toffoli Gates with Intermediate Qudit}
\begin{figure}[!h]
            \centering
            \begin{quantikz}
                \lstick{\ket{q_0}}&\ctrl{1}&\\
                \lstick{\ket{q_1}}&\ctrl{1}&\\
                \lstick{\ket{q_2}}&\targ{}&
            \end{quantikz}
            $\equiv$
            \begin{quantikz}
                \lstick{\ket{q_0}}&\measure{1}\highlightCircuitTopLabel{2}{1}{blue}{Preparation}& &\measure{1}\highlightCircuitTopLabel{2}{1}{red}{Correction}& \\
                \lstick{\ket{q_1}}&\gate{X_{3}^{+1}}\wire[u]{a}&\measure{2}\wire[d]{a} \highlightCircuitBottomLabel{2}{1}{green}{Target Operation}&\gate{X_{3}^{-1}}\wire[u]{a}& \\
                \lstick{\ket{q_2}}& &\targ{}& &
            \end{quantikz}
            \caption{2-controlled Toffoli decomposition using intermediate qutrits.}
            \label{fig:2-control-toffoli-decomposition}
        \end{figure}
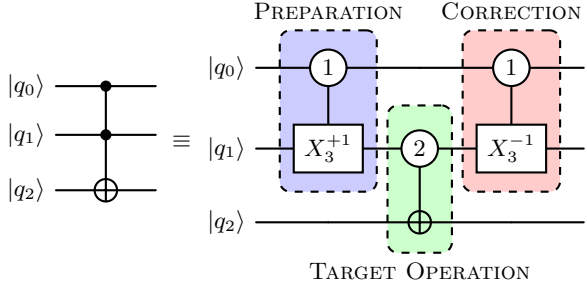

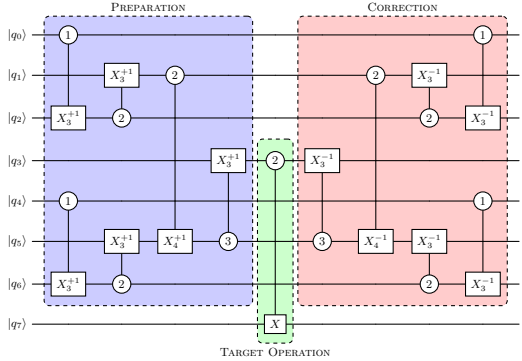
\begin{figure}[!h]
            \centering
            \scalebox{0.5}{
\begin{quantikz}
                \lstick{\ket{q_0}} &\measure{1}\wire[d]{a}\highlightCircuitTopLabel{7}{4}{blue}{Preparation}& & & & &\highlightCircuitTopLabel{7}{4}{red}{Correction}& & &\measure{1}\wire[d]{a}& \\
                \lstick{\ket{q_1}} &\wire[d]{a}&\gate{X^{+1}_3}&\measure{2}\wire[d]{a}& & & &\measure{2}\wire[d]{a}&\gate{X^{-1}_3}&\wire[d]{a}& \\
                \lstick{\ket{q_2}} &\gate{X^{+1}_3}&\measure{2}\wire[u]{a}&\wire[d]{a}& & & &\wire[d]{a}&\measure{2}\wire[u]{a}&\gate{X^{-1}_3}& \\
                \lstick{\ket{q_3}} & & &\wire[d]{a}&\gate{X^{+1}_3}\wire[d]{a}&\measure{2}\wire[d]{a}\highlightCircuitBottomLabel{5}{1}{green}{Target Operation}&\gate{X^{-1}_3}\wire[d]{a}&\wire[d]{a}&&& \\
                \lstick{\ket{q_4}} &\measure{1}\wire[d]{a}& &\wire[d]{a}&\wire[d]{a}&\wire[d]{a}&\wire[d]{a}&\wire[d]{a}& &\measure{1}\wire[d]{a}& \\
                \lstick{\ket{q_5}} &\wire[d]{a}&\gate{X^{+1}_3}&\gate{X^{+1}_4}&\measure{3}&\wire[d]{a}&\measure{3}&\gate{X^{-1}_4}&\gate{X^{-1}_3}&\wire[d]{a}& \\
                \lstick{\ket{q_6}} &\gate{X^{+1}_3}&\measure{2}\wire[u]{a}& & &\wire[d]{a}& & &\measure{2}\wire[u]{a}&\gate{X^{-1}_3}& \\
                \lstick{\ket{q_7}} & & & & &\gate{X}& & &&&
            \end{quantikz}}
            \caption{7-controlled Toffoli decomposition using intermediate-qutrit and intermediate-ququad.}
\label{8-qubit}
    \end{figure}

    To achieve more improvement, we use multi-controlled Toffoli decomposition using intermediate qudits \cite{sahaintermediate, toffoli_decomposition_gokhale} in our proposed quantum walk circuit, where momentarily the binary system goes into higher dimensions, and using this concept, after the decomposition, gate count and circuit depth are reduced. No ancilla bits are required for this decomposition. In \cite{toffoli_decomposition_gokhale}, the authors have shown how we can efficiently decompose a multi-controlled Toffoli gate in the binary system by employing qutrits (3-ary quantum systems). In \cite{sahaintermediate}, the authors have extended and generalized the previous work to decompose $n$-qubit Toffoli gates by considering both the intermediate-qutrit and the intermediate-ququad (4-ary quantum systems). For instance, in Figure \ref{fig:2-control-toffoli-decomposition}, a 2-controlled Toffoli gate is decomposed into 2-qutrit gates. We have also portrayed 7-controlled Toffloi gate decomposition into 2-qutrit gates and 2-ququad gates in Figure \ref{8-qubit}.

    For the sake of better understandability, we have divided the steps of decomposition into three steps - 
        \begin{itemize}
            \item \textbf{Preparation:} In this step, the controls are prepared for the operation on the target. We must use only 2-qudit gates, and in the original circuit, we had multiple controls (more than 1). Therefore, based on the state of all those controls, we need to manipulate a single control in such a way that if all the other controls are set to $\ket{1}$, then only the last single control will be set to $\ket{2}$. However, in the process of setting a single control to $\ket{2}$, the state of the other controls may get disturbed. Thus, in the final step, we need to correct those disturbances. We have highlighted this step with purple color.
            \item \textbf{Target Operation:} Following the initial step, where a single control is set to $\ket{2}$ if all other controls were $\ket{1}$, this step involves simply applying a CNOT gate. The control for this gate is the single control that was set to $\ket{2}$ in the previous step, while the target remains the original target. We have highlighted this step with green color.
            \item \textbf{Correction:} As previously discussed, this final step aims to correct the states of the controls. We need to reverse the operation performed in the first step, ensuring that only the target changes if all the controls were set to $\ket{1}$. We have highlighted this step with red color.
        \end{itemize}
        Using this logic, we can decompose binary multi-controlled Toffoli gates of any size using intermediate qudits. This can be generalized for any finite $d$-dimensional quantum system as per \cite{sahaintermediate} using two higher dimensions i.e., \ket{d+1} and \ket{d+2}, which is thoroughly discussed in Appendix \ref{appendix: intermediate}.

\begin{figure}[!h]
            \centering
            \includegraphics[scale=0.5]{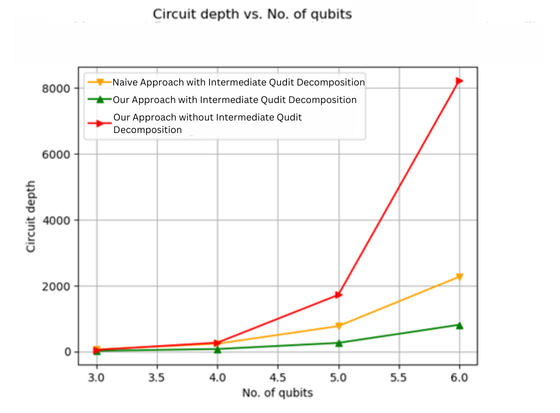}
            \caption{Comparative analysis of circuit depth.}
            \label{fig:cir_depth_analysis_intermediate}
        \end{figure}

    We further observe that by using intermediate qudit decomposition, we have acutely reduced the gate count and circuit depth. For the $N$-controlled Toffoli gate, we get a circuit depth of $\log{N}$ \cite{sahaintermediate}. By considering that we can formulate a generalized formulation for our proposed quantum walk circuit. Thus, the total circuit depth for a $N$ qubit circuit is $2^{N-1}(\sum_{x=2}^{N-1} (\log{x}) + 1) + 3*2^{N-2} + 2^{N-2}$.
        Similarly the total gate count for a $N$ qubit circuit is $2^{N-1}(\sum_{x=2}^{N-1} (2x-3) + 1) + 2*2^{N-2} + 4*2^{N-2}$.   We have compared circuit depth among naive approach \cite{saha1} with intermediate qudits, enhanced increment decrement with intermediate qudits, and increment decrement without using intermediate qudits. The comparative analysis is shown in Figure \ref{fig:cir_depth_analysis_intermediate}.  

\begin{figure}[!h]
        \centering
        \includegraphics[width=0.42\textwidth]{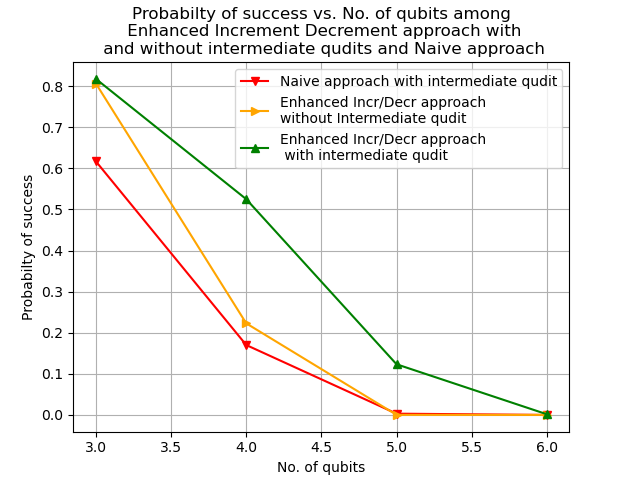}
        \caption{Comparative analysis of success probability.}
        \label{fig:prob_success_analysis_intermediate}
    \end{figure}

       With these circuit costs, we have also compared the probability of success by considering Equation \ref{prob_success} among the naive approach, enhanced increment decrement approach using intermediate qudits, and increment decrement approach without using intermediate qudits. The comparative analysis is shown in Figure \ref{fig:prob_success_analysis_intermediate}. Using intermediate qudit decomposition for our approach increases the probability of success. We have also simulated and validated our proposed quantum walk circuits with intermediate qudits on the QuDiet platform \cite{Chatterjee_2023}. 

\section{Conclusion}\label{conclusion}
This paper achieved a 50\% reduction in circuit cost, including gate count and circuit depth, compared to the current increment-decrement method, through our proposed technique in qubit systems. The experimental results extended the significance beyond a few time steps, establishing a reliable foundation for quantum computer utilization. Furthermore, the engineering excellence of our proposed approach achieves the generalization for any finite-dimensional quantum system. Additionally, to ensure efficient implementation of quantum walks without ancillary requirements, we integrated an intermediate qudit technique for multi-qubit gate decomposition. This proposed approach can be easily employed for higher dimensional lattices as well. How this proposed circuit can enhance the applications of quantum walks \cite{Srikara_2020, Saha_2022} remains a prospect outlined in this paper. The code for our proposed implementation of DTQW is available in \href{https://github.com/Biswayan375/Discrete-Time-Quantum-Walk/tree/enh-enc-dec}{https://github.com/Biswayan375/Discrete-Time-Quantum-Walk/tree/enh-enc-dec}. 

\bibliography{sample-base}

\begin{thebibliography}{49}%
\makeatletter
\providecommand \@ifxundefined [1]{%
 \@ifx{#1\undefined}
}%
\providecommand \@ifnum [1]{%
 \ifnum #1\expandafter \@firstoftwo
 \else \expandafter \@secondoftwo
 \fi
}%
\providecommand \@ifx [1]{%
 \ifx #1\expandafter \@firstoftwo
 \else \expandafter \@secondoftwo
 \fi
}%
\providecommand \natexlab [1]{#1}%
\providecommand \enquote  [1]{``#1''}%
\providecommand \bibnamefont  [1]{#1}%
\providecommand \bibfnamefont [1]{#1}%
\providecommand \citenamefont [1]{#1}%
\providecommand \href@noop [0]{\@secondoftwo}%
\providecommand \href [0]{\begingroup \@sanitize@url \@href}%
\providecommand \@href[1]{\@@startlink{#1}\@@href}%
\providecommand \@@href[1]{\endgroup#1\@@endlink}%
\providecommand \@sanitize@url [0]{\catcode `\\12\catcode `\$12\catcode `\&12\catcode `\#12\catcode `\^12\catcode `\_12\catcode `\%12\relax}%
\providecommand \@@startlink[1]{}%
\providecommand \@@endlink[0]{}%
\providecommand \url  [0]{\begingroup\@sanitize@url \@url }%
\providecommand \@url [1]{\endgroup\@href {#1}{\urlprefix }}%
\providecommand \urlprefix  [0]{URL }%
\providecommand \Eprint [0]{\href }%
\providecommand \doibase [0]{https://doi.org/}%
\providecommand \selectlanguage [0]{\@gobble}%
\providecommand \bibinfo  [0]{\@secondoftwo}%
\providecommand \bibfield  [0]{\@secondoftwo}%
\providecommand \translation [1]{[#1]}%
\providecommand \BibitemOpen [0]{}%
\providecommand \bibitemStop [0]{}%
\providecommand \bibitemNoStop [0]{.\EOS\space}%
\providecommand \EOS [0]{\spacefactor3000\relax}%
\providecommand \BibitemShut  [1]{\csname bibitem#1\endcsname}%
\let\auto@bib@innerbib\@empty
\bibitem [{\citenamefont {Feynman}(2018)}]{feynman2018simulating}%
  \BibitemOpen
  \bibfield  {author} {\bibinfo {author} {\bibfnamefont {R.~P.}\ \bibnamefont {Feynman}},\ }\bibfield  {title} {\bibinfo {title} {Simulating physics with computers},\ }in\ \href@noop {} {\emph {\bibinfo {booktitle} {Feynman and computation}}}\ (\bibinfo  {publisher} {CRC Press},\ \bibinfo {year} {2018})\ pp.\ \bibinfo {pages} {133--153}\BibitemShut {NoStop}%
\bibitem [{\citenamefont {Nielsen}\ and\ \citenamefont {Chuang}(2010)}]{nielsenchuang}%
  \BibitemOpen
  \bibfield  {author} {\bibinfo {author} {\bibfnamefont {M.~A.}\ \bibnamefont {Nielsen}}\ and\ \bibinfo {author} {\bibfnamefont {I.~L.}\ \bibnamefont {Chuang}},\ }\href@noop {} {\emph {\bibinfo {title} {Quantum computation and quantum information}}},\ Vol.\ \bibinfo {volume} {10th Anniversary Edition}\ (\bibinfo  {publisher} {Cambridge University Press},\ \bibinfo {year} {2010})\BibitemShut {NoStop}%
\bibitem [{\citenamefont {Grover}(1996)}]{grover1996fast}%
  \BibitemOpen
  \bibfield  {author} {\bibinfo {author} {\bibfnamefont {L.~K.}\ \bibnamefont {Grover}},\ }\bibfield  {title} {\bibinfo {title} {A fast quantum mechanical algorithm for database search},\ }in\ \href@noop {} {\emph {\bibinfo {booktitle} {Proceedings of the twenty-eighth annual ACM symposium on Theory of computing}}}\ (\bibinfo {year} {1996})\ pp.\ \bibinfo {pages} {212--219}\BibitemShut {NoStop}%
\bibitem [{\citenamefont {Shor}(1994)}]{shor1994algorithms}%
  \BibitemOpen
  \bibfield  {author} {\bibinfo {author} {\bibfnamefont {P.~W.}\ \bibnamefont {Shor}},\ }\bibfield  {title} {\bibinfo {title} {Algorithms for quantum computation: discrete logarithms and factoring},\ }in\ \href@noop {} {\emph {\bibinfo {booktitle} {Proceedings 35th annual symposium on foundations of computer science}}}\ (\bibinfo {organization} {Ieee},\ \bibinfo {year} {1994})\ pp.\ \bibinfo {pages} {124--134}\BibitemShut {NoStop}%
\bibitem [{\citenamefont {Aharonov}\ \emph {et~al.}(1993)\citenamefont {Aharonov}, \citenamefont {Davidovich},\ and\ \citenamefont {Zagury}}]{aharonov1993quantum}%
  \BibitemOpen
  \bibfield  {author} {\bibinfo {author} {\bibfnamefont {Y.}~\bibnamefont {Aharonov}}, \bibinfo {author} {\bibfnamefont {L.}~\bibnamefont {Davidovich}},\ and\ \bibinfo {author} {\bibfnamefont {N.}~\bibnamefont {Zagury}},\ }\bibfield  {title} {\bibinfo {title} {Quantum random walks},\ }\href@noop {} {\bibfield  {journal} {\bibinfo  {journal} {Physical Review A}\ }\textbf {\bibinfo {volume} {48}},\ \bibinfo {pages} {1687} (\bibinfo {year} {1993})}\BibitemShut {NoStop}%
\bibitem [{\citenamefont {Kac}(1947)}]{brownian1947}%
  \BibitemOpen
  \bibfield  {author} {\bibinfo {author} {\bibfnamefont {M.}~\bibnamefont {Kac}},\ }\bibfield  {title} {\bibinfo {title} {Random walk and the theory of brownian motion},\ }\href@noop {} {\bibfield  {journal} {\bibinfo  {journal} {The American Mathematical Monthly}\ }\textbf {\bibinfo {volume} {54}},\ \bibinfo {pages} {369} (\bibinfo {year} {1947})}\BibitemShut {NoStop}%
\bibitem [{\citenamefont {Godfrey}\ \emph {et~al.}(1964)\citenamefont {Godfrey}, \citenamefont {Granger},\ and\ \citenamefont {Morgenstern}}]{stockprice1964}%
  \BibitemOpen
  \bibfield  {author} {\bibinfo {author} {\bibfnamefont {M.~D.}\ \bibnamefont {Godfrey}}, \bibinfo {author} {\bibfnamefont {C.~W.}\ \bibnamefont {Granger}},\ and\ \bibinfo {author} {\bibfnamefont {O.}~\bibnamefont {Morgenstern}},\ }\bibfield  {title} {\bibinfo {title} {The random-walk hypothesis of stock market behavior a},\ }\href@noop {} {\bibfield  {journal} {\bibinfo  {journal} {Kyklos}\ }\textbf {\bibinfo {volume} {17}},\ \bibinfo {pages} {1} (\bibinfo {year} {1964})}\BibitemShut {NoStop}%
\bibitem [{\citenamefont {Page}\ \emph {et~al.}(1998)\citenamefont {Page}, \citenamefont {Brin}, \citenamefont {Motwani},\ and\ \citenamefont {Winograd}}]{pagerank}%
  \BibitemOpen
  \bibfield  {author} {\bibinfo {author} {\bibfnamefont {L.}~\bibnamefont {Page}}, \bibinfo {author} {\bibfnamefont {S.}~\bibnamefont {Brin}}, \bibinfo {author} {\bibfnamefont {R.}~\bibnamefont {Motwani}},\ and\ \bibinfo {author} {\bibfnamefont {T.}~\bibnamefont {Winograd}},\ }\href@noop {} {\emph {\bibinfo {title} {The pagerank citation ranking: Bring order to the web}}},\ \bibinfo {type} {Tech. Rep.}\ (\bibinfo  {institution} {Technical report, Stanford University},\ \bibinfo {year} {1998})\BibitemShut {NoStop}%
\bibitem [{\citenamefont {Kempe}(2003)}]{kempe2003quantum}%
  \BibitemOpen
  \bibfield  {author} {\bibinfo {author} {\bibfnamefont {J.}~\bibnamefont {Kempe}},\ }\bibfield  {title} {\bibinfo {title} {Quantum random walks: an introductory overview},\ }\href@noop {} {\bibfield  {journal} {\bibinfo  {journal} {Contemporary Physics}\ }\textbf {\bibinfo {volume} {44}},\ \bibinfo {pages} {307} (\bibinfo {year} {2003})}\BibitemShut {NoStop}%
\bibitem [{\citenamefont {Abd El-Latif}\ \emph {et~al.}(2020)\citenamefont {Abd El-Latif}, \citenamefont {Abd-El-Atty}, \citenamefont {Amin},\ and\ \citenamefont {Iliyasu}}]{crypto2020}%
  \BibitemOpen
  \bibfield  {author} {\bibinfo {author} {\bibfnamefont {A.~A.}\ \bibnamefont {Abd El-Latif}}, \bibinfo {author} {\bibfnamefont {B.}~\bibnamefont {Abd-El-Atty}}, \bibinfo {author} {\bibfnamefont {M.}~\bibnamefont {Amin}},\ and\ \bibinfo {author} {\bibfnamefont {A.~M.}\ \bibnamefont {Iliyasu}},\ }\bibfield  {title} {\bibinfo {title} {Quantum-inspired cascaded discrete-time quantum walks with induced chaotic dynamics and cryptographic applications},\ }\href@noop {} {\bibfield  {journal} {\bibinfo  {journal} {Scientific reports}\ }\textbf {\bibinfo {volume} {10}},\ \bibinfo {pages} {1930} (\bibinfo {year} {2020})}\BibitemShut {NoStop}%
\bibitem [{\citenamefont {Feynman}\ \emph {et~al.}(2010)\citenamefont {Feynman}, \citenamefont {Hibbs},\ and\ \citenamefont {Styer}}]{feynman2010quantum}%
  \BibitemOpen
  \bibfield  {author} {\bibinfo {author} {\bibfnamefont {R.~P.}\ \bibnamefont {Feynman}}, \bibinfo {author} {\bibfnamefont {A.~R.}\ \bibnamefont {Hibbs}},\ and\ \bibinfo {author} {\bibfnamefont {D.~F.}\ \bibnamefont {Styer}},\ }\href@noop {} {\emph {\bibinfo {title} {Quantum mechanics and path integrals}}}\ (\bibinfo  {publisher} {Courier Corporation},\ \bibinfo {year} {2010})\BibitemShut {NoStop}%
\bibitem [{\citenamefont {Farhi}\ and\ \citenamefont {Gutmann}(1998)}]{farhi1998quantum}%
  \BibitemOpen
  \bibfield  {author} {\bibinfo {author} {\bibfnamefont {E.}~\bibnamefont {Farhi}}\ and\ \bibinfo {author} {\bibfnamefont {S.}~\bibnamefont {Gutmann}},\ }\bibfield  {title} {\bibinfo {title} {Quantum computation and decision trees},\ }\href@noop {} {\bibfield  {journal} {\bibinfo  {journal} {Physical Review A}\ }\textbf {\bibinfo {volume} {58}},\ \bibinfo {pages} {915} (\bibinfo {year} {1998})}\BibitemShut {NoStop}%
\bibitem [{\citenamefont {Childs}(2009)}]{childs2009universal}%
  \BibitemOpen
  \bibfield  {author} {\bibinfo {author} {\bibfnamefont {A.~M.}\ \bibnamefont {Childs}},\ }\bibfield  {title} {\bibinfo {title} {Universal computation by quantum walk},\ }\href@noop {} {\bibfield  {journal} {\bibinfo  {journal} {Physical review letters}\ }\textbf {\bibinfo {volume} {102}},\ \bibinfo {pages} {180501} (\bibinfo {year} {2009})}\BibitemShut {NoStop}%
\bibitem [{\citenamefont {Childs}\ \emph {et~al.}(2013)\citenamefont {Childs}, \citenamefont {Gosset},\ and\ \citenamefont {Webb}}]{childs2013universal}%
  \BibitemOpen
  \bibfield  {author} {\bibinfo {author} {\bibfnamefont {A.~M.}\ \bibnamefont {Childs}}, \bibinfo {author} {\bibfnamefont {D.}~\bibnamefont {Gosset}},\ and\ \bibinfo {author} {\bibfnamefont {Z.}~\bibnamefont {Webb}},\ }\bibfield  {title} {\bibinfo {title} {Universal computation by multiparticle quantum walk},\ }\href@noop {} {\bibfield  {journal} {\bibinfo  {journal} {Science}\ }\textbf {\bibinfo {volume} {339}},\ \bibinfo {pages} {791} (\bibinfo {year} {2013})}\BibitemShut {NoStop}%
\bibitem [{\citenamefont {Ambainis}(2007)}]{ambainis2007quantum}%
  \BibitemOpen
  \bibfield  {author} {\bibinfo {author} {\bibfnamefont {A.}~\bibnamefont {Ambainis}},\ }\bibfield  {title} {\bibinfo {title} {Quantum walk algorithm for element distinctness},\ }\href@noop {} {\bibfield  {journal} {\bibinfo  {journal} {SIAM Journal on Computing}\ }\textbf {\bibinfo {volume} {37}},\ \bibinfo {pages} {210} (\bibinfo {year} {2007})}\BibitemShut {NoStop}%
\bibitem [{\citenamefont {Engel}\ \emph {et~al.}(2007)\citenamefont {Engel}, \citenamefont {Calhoun}, \citenamefont {Read}, \citenamefont {Ahn}, \citenamefont {Man{\v{c}}al}, \citenamefont {Cheng}, \citenamefont {Blankenship},\ and\ \citenamefont {Fleming}}]{engel2007evidence}%
  \BibitemOpen
  \bibfield  {author} {\bibinfo {author} {\bibfnamefont {G.~S.}\ \bibnamefont {Engel}}, \bibinfo {author} {\bibfnamefont {T.~R.}\ \bibnamefont {Calhoun}}, \bibinfo {author} {\bibfnamefont {E.~L.}\ \bibnamefont {Read}}, \bibinfo {author} {\bibfnamefont {T.-K.}\ \bibnamefont {Ahn}}, \bibinfo {author} {\bibfnamefont {T.}~\bibnamefont {Man{\v{c}}al}}, \bibinfo {author} {\bibfnamefont {Y.-C.}\ \bibnamefont {Cheng}}, \bibinfo {author} {\bibfnamefont {R.~E.}\ \bibnamefont {Blankenship}},\ and\ \bibinfo {author} {\bibfnamefont {G.~R.}\ \bibnamefont {Fleming}},\ }\bibfield  {title} {\bibinfo {title} {Evidence for wavelike energy transfer through quantum coherence in photosynthetic systems},\ }\href@noop {} {\bibfield  {journal} {\bibinfo  {journal} {Nature}\ }\textbf {\bibinfo {volume} {446}},\ \bibinfo {pages} {782} (\bibinfo {year} {2007})}\BibitemShut {NoStop}%
\bibitem [{\citenamefont {Kendon}(2006)}]{kendon2006random}%
  \BibitemOpen
  \bibfield  {author} {\bibinfo {author} {\bibfnamefont {V.~M.}\ \bibnamefont {Kendon}},\ }\bibfield  {title} {\bibinfo {title} {A random walk approach to quantum algorithms},\ }\href@noop {} {\bibfield  {journal} {\bibinfo  {journal} {Philosophical Transactions of the Royal Society A: Mathematical, Physical and Engineering Sciences}\ }\textbf {\bibinfo {volume} {364}},\ \bibinfo {pages} {3407} (\bibinfo {year} {2006})}\BibitemShut {NoStop}%
\bibitem [{\citenamefont {Wadhia}\ \emph {et~al.}(2024)\citenamefont {Wadhia}, \citenamefont {Chancellor},\ and\ \citenamefont {Kendon}}]{wadhia2024cycle}%
  \BibitemOpen
  \bibfield  {author} {\bibinfo {author} {\bibfnamefont {V.}~\bibnamefont {Wadhia}}, \bibinfo {author} {\bibfnamefont {N.}~\bibnamefont {Chancellor}},\ and\ \bibinfo {author} {\bibfnamefont {V.}~\bibnamefont {Kendon}},\ }\bibfield  {title} {\bibinfo {title} {Cycle discrete-time quantum walks on a noisy quantum computer},\ }\href@noop {} {\bibfield  {journal} {\bibinfo  {journal} {The European Physical Journal D}\ }\textbf {\bibinfo {volume} {78}},\ \bibinfo {pages} {29} (\bibinfo {year} {2024})}\BibitemShut {NoStop}%
\bibitem [{\citenamefont {Douglas}\ and\ \citenamefont {Wang}(2009{\natexlab{a}})}]{douglas2009efficient}%
  \BibitemOpen
  \bibfield  {author} {\bibinfo {author} {\bibfnamefont {B.}~\bibnamefont {Douglas}}\ and\ \bibinfo {author} {\bibfnamefont {J.}~\bibnamefont {Wang}},\ }\bibfield  {title} {\bibinfo {title} {Efficient quantum circuit implementation of quantum walks},\ }\href@noop {} {\bibfield  {journal} {\bibinfo  {journal} {Physical Review A}\ }\textbf {\bibinfo {volume} {79}},\ \bibinfo {pages} {052335} (\bibinfo {year} {2009}{\natexlab{a}})}\BibitemShut {NoStop}%
\bibitem [{\citenamefont {Douglas}\ and\ \citenamefont {Wang}(2009{\natexlab{b}})}]{ambainis2001one}%
  \BibitemOpen
  \bibfield  {author} {\bibinfo {author} {\bibfnamefont {B.~L.}\ \bibnamefont {Douglas}}\ and\ \bibinfo {author} {\bibfnamefont {J.~B.}\ \bibnamefont {Wang}},\ }\bibfield  {title} {\bibinfo {title} {Efficient quantum circuit implementation of quantum walks},\ }\href {https://doi.org/10.1103/PhysRevA.79.052335} {\bibfield  {journal} {\bibinfo  {journal} {Phys. Rev. A}\ }\textbf {\bibinfo {volume} {79}},\ \bibinfo {pages} {052335} (\bibinfo {year} {2009}{\natexlab{b}})}\BibitemShut {NoStop}%
\bibitem [{\citenamefont {Singh}\ \emph {et~al.}(2021{\natexlab{a}})\citenamefont {Singh}, \citenamefont {Alderete}, \citenamefont {Balu}, \citenamefont {Monroe}, \citenamefont {Linke},\ and\ \citenamefont {Chandrashekar}}]{singh2021quantum}%
  \BibitemOpen
  \bibfield  {author} {\bibinfo {author} {\bibfnamefont {S.}~\bibnamefont {Singh}}, \bibinfo {author} {\bibfnamefont {C.~H.}\ \bibnamefont {Alderete}}, \bibinfo {author} {\bibfnamefont {R.}~\bibnamefont {Balu}}, \bibinfo {author} {\bibfnamefont {C.}~\bibnamefont {Monroe}}, \bibinfo {author} {\bibfnamefont {N.~M.}\ \bibnamefont {Linke}},\ and\ \bibinfo {author} {\bibfnamefont {C.}~\bibnamefont {Chandrashekar}},\ }\bibfield  {title} {\bibinfo {title} {Quantum circuits for the realization of equivalent forms of one-dimensional discrete-time quantum walks on near-term quantum hardware},\ }\href@noop {} {\bibfield  {journal} {\bibinfo  {journal} {Physical Review A}\ }\textbf {\bibinfo {volume} {104}},\ \bibinfo {pages} {062401} (\bibinfo {year} {2021}{\natexlab{a}})}\BibitemShut {NoStop}%
\bibitem [{\citenamefont {Razzoli}\ \emph {et~al.}(2024)\citenamefont {Razzoli}, \citenamefont {Cenedese}, \citenamefont {Bondani},\ and\ \citenamefont {Benenti}}]{Razzoli_2024}%
  \BibitemOpen
  \bibfield  {author} {\bibinfo {author} {\bibfnamefont {L.}~\bibnamefont {Razzoli}}, \bibinfo {author} {\bibfnamefont {G.}~\bibnamefont {Cenedese}}, \bibinfo {author} {\bibfnamefont {M.}~\bibnamefont {Bondani}},\ and\ \bibinfo {author} {\bibfnamefont {G.}~\bibnamefont {Benenti}},\ }\bibfield  {title} {\bibinfo {title} {Efficient implementation of discrete-time quantum walks on quantum computers},\ }\href {https://doi.org/10.3390/e26040313} {\bibfield  {journal} {\bibinfo  {journal} {Entropy}\ }\textbf {\bibinfo {volume} {26}},\ \bibinfo {pages} {313} (\bibinfo {year} {2024})}\BibitemShut {NoStop}%
\bibitem [{\citenamefont {Preskill}(2018)}]{Preskill_2018}%
  \BibitemOpen
  \bibfield  {author} {\bibinfo {author} {\bibfnamefont {J.}~\bibnamefont {Preskill}},\ }\bibfield  {title} {\bibinfo {title} {Quantum computing in the nisq era and beyond},\ }\href {https://doi.org/10.22331/q-2018-08-06-79} {\bibfield  {journal} {\bibinfo  {journal} {Quantum}\ }\textbf {\bibinfo {volume} {2}},\ \bibinfo {pages} {79} (\bibinfo {year} {2018})}\BibitemShut {NoStop}%
\bibitem [{\citenamefont {Saha}\ \emph {et~al.}(2021)\citenamefont {Saha}, \citenamefont {Mandal}, \citenamefont {Saha},\ and\ \citenamefont {Chakrabarti}}]{9410395}%
  \BibitemOpen
  \bibfield  {author} {\bibinfo {author} {\bibfnamefont {A.}~\bibnamefont {Saha}}, \bibinfo {author} {\bibfnamefont {S.~B.}\ \bibnamefont {Mandal}}, \bibinfo {author} {\bibfnamefont {D.}~\bibnamefont {Saha}},\ and\ \bibinfo {author} {\bibfnamefont {A.}~\bibnamefont {Chakrabarti}},\ }\bibfield  {title} {\bibinfo {title} {One-dimensional lazy quantum walk in ternary system},\ }\href {https://doi.org/10.1109/TQE.2021.3074707} {\bibfield  {journal} {\bibinfo  {journal} {IEEE Transactions on Quantum Engineering}\ }\textbf {\bibinfo {volume} {2}},\ \bibinfo {pages} {1} (\bibinfo {year} {2021})}\BibitemShut {NoStop}%
\bibitem [{\citenamefont {Saha}\ \emph {et~al.}(2022{\natexlab{a}})\citenamefont {Saha}, \citenamefont {Saha},\ and\ \citenamefont {Chakrabarti}}]{saha1}%
  \BibitemOpen
  \bibfield  {author} {\bibinfo {author} {\bibfnamefont {A.}~\bibnamefont {Saha}}, \bibinfo {author} {\bibfnamefont {D.}~\bibnamefont {Saha}},\ and\ \bibinfo {author} {\bibfnamefont {A.}~\bibnamefont {Chakrabarti}},\ }\bibfield  {title} {\bibinfo {title} {Discrete-time quantum walks in qudit systems},\ }\href@noop {} {\bibfield  {journal} {\bibinfo  {journal} {arXiv preprint arXiv:2207.04319}\ } (\bibinfo {year} {2022}{\natexlab{a}})}\BibitemShut {NoStop}%
\bibitem [{\citenamefont {Jayakody}\ \emph {et~al.}(2021)\citenamefont {Jayakody}, \citenamefont {Meena},\ and\ \citenamefont {Pradhan}}]{jayakody}%
  \BibitemOpen
  \bibfield  {author} {\bibinfo {author} {\bibfnamefont {M.~N.}\ \bibnamefont {Jayakody}}, \bibinfo {author} {\bibfnamefont {C.}~\bibnamefont {Meena}},\ and\ \bibinfo {author} {\bibfnamefont {P.}~\bibnamefont {Pradhan}},\ }\bibfield  {title} {\bibinfo {title} {One-dimensional discrete-time quantum walks with general coin},\ }\href@noop {} {\bibfield  {journal} {\bibinfo  {journal} {arXiv preprint arXiv:2102.07207}\ } (\bibinfo {year} {2021})}\BibitemShut {NoStop}%
\bibitem [{\citenamefont {Singh}\ \emph {et~al.}(2021{\natexlab{b}})\citenamefont {Singh}, \citenamefont {Chawla}, \citenamefont {Sarkar},\ and\ \citenamefont {Chandrashekar}}]{shivanisingh}%
  \BibitemOpen
  \bibfield  {author} {\bibinfo {author} {\bibfnamefont {S.}~\bibnamefont {Singh}}, \bibinfo {author} {\bibfnamefont {P.}~\bibnamefont {Chawla}}, \bibinfo {author} {\bibfnamefont {A.}~\bibnamefont {Sarkar}},\ and\ \bibinfo {author} {\bibfnamefont {C.}~\bibnamefont {Chandrashekar}},\ }\bibfield  {title} {\bibinfo {title} {Universal quantum computing using single-particle discrete-time quantum walk},\ }\href@noop {} {\bibfield  {journal} {\bibinfo  {journal} {Scientific Reports}\ }\textbf {\bibinfo {volume} {11}},\ \bibinfo {pages} {11551} (\bibinfo {year} {2021}{\natexlab{b}})}\BibitemShut {NoStop}%
\bibitem [{\citenamefont {Saha}\ \emph {et~al.}(2022{\natexlab{b}})\citenamefont {Saha}, \citenamefont {Majumdar}, \citenamefont {Saha}, \citenamefont {Chakrabarti},\ and\ \citenamefont {Sur-Kolay}}]{sahaintermediate}%
  \BibitemOpen
  \bibfield  {author} {\bibinfo {author} {\bibfnamefont {A.}~\bibnamefont {Saha}}, \bibinfo {author} {\bibfnamefont {R.}~\bibnamefont {Majumdar}}, \bibinfo {author} {\bibfnamefont {D.}~\bibnamefont {Saha}}, \bibinfo {author} {\bibfnamefont {A.}~\bibnamefont {Chakrabarti}},\ and\ \bibinfo {author} {\bibfnamefont {S.}~\bibnamefont {Sur-Kolay}},\ }\bibfield  {title} {\bibinfo {title} {Asymptotically improved circuit for a d-ary grover's algorithm with advanced decomposition of the n-qudit toffoli gate},\ }\href@noop {} {\bibfield  {journal} {\bibinfo  {journal} {Physical Review A}\ }\textbf {\bibinfo {volume} {105}},\ \bibinfo {pages} {062453} (\bibinfo {year} {2022}{\natexlab{b}})}\BibitemShut {NoStop}%
\bibitem [{\citenamefont {Majumdar}\ \emph {et~al.}(2024)\citenamefont {Majumdar}, \citenamefont {Madan}, \citenamefont {Bhoumik}, \citenamefont {Vinayagamurthy}, \citenamefont {Raghunathan},\ and\ \citenamefont {Sur-Kolay}}]{majumdar2021optimizing}%
  \BibitemOpen
  \bibfield  {author} {\bibinfo {author} {\bibfnamefont {R.}~\bibnamefont {Majumdar}}, \bibinfo {author} {\bibfnamefont {D.}~\bibnamefont {Madan}}, \bibinfo {author} {\bibfnamefont {D.}~\bibnamefont {Bhoumik}}, \bibinfo {author} {\bibfnamefont {D.}~\bibnamefont {Vinayagamurthy}}, \bibinfo {author} {\bibfnamefont {S.}~\bibnamefont {Raghunathan}},\ and\ \bibinfo {author} {\bibfnamefont {S.}~\bibnamefont {Sur-Kolay}},\ }\bibfield  {title} {\bibinfo {title} {Optimized qaoa ansatz circuit design for two-body hamiltonian problems},\ }in\ \href {https://doi.org/10.1109/VLSID60093.2024.00072} {\emph {\bibinfo {booktitle} {2024 37th International Conference on VLSI Design and 2024 23rd International Conference on Embedded Systems (VLSID)}}}\ (\bibinfo {year} {2024})\ pp.\ \bibinfo {pages} {396--401}\BibitemShut {NoStop}%
\bibitem [{\citenamefont {Fischer}\ \emph {et~al.}(2022)\citenamefont {Fischer}, \citenamefont {Miller}, \citenamefont {Tacchino}, \citenamefont {Barkoutsos}, \citenamefont {Egger},\ and\ \citenamefont {Tavernelli}}]{https://doi.org/10.48550/arxiv.2203.07369}%
  \BibitemOpen
  \bibfield  {author} {\bibinfo {author} {\bibfnamefont {L.~E.}\ \bibnamefont {Fischer}}, \bibinfo {author} {\bibfnamefont {D.}~\bibnamefont {Miller}}, \bibinfo {author} {\bibfnamefont {F.}~\bibnamefont {Tacchino}}, \bibinfo {author} {\bibfnamefont {P.~K.}\ \bibnamefont {Barkoutsos}}, \bibinfo {author} {\bibfnamefont {D.~J.}\ \bibnamefont {Egger}},\ and\ \bibinfo {author} {\bibfnamefont {I.}~\bibnamefont {Tavernelli}},\ }\bibfield  {title} {\bibinfo {title} {Ancilla-free implementation of generalized measurements for qubits embedded in a qudit space},\ }\href {https://doi.org/10.1103/PhysRevResearch.4.033027} {\bibfield  {journal} {\bibinfo  {journal} {Phys. Rev. Res.}\ }\textbf {\bibinfo {volume} {4}},\ \bibinfo {pages} {033027} (\bibinfo {year} {2022})}\BibitemShut {NoStop}%
\bibitem [{\citenamefont {Saeedi}\ and\ \citenamefont {Pedram}(2013)}]{PhysRevA.87.062318}%
  \BibitemOpen
  \bibfield  {author} {\bibinfo {author} {\bibfnamefont {M.}~\bibnamefont {Saeedi}}\ and\ \bibinfo {author} {\bibfnamefont {M.}~\bibnamefont {Pedram}},\ }\bibfield  {title} {\bibinfo {title} {Linear-depth quantum circuits for $n$-qubit toffoli gates with no ancilla},\ }\href {https://doi.org/10.1103/PhysRevA.87.062318} {\bibfield  {journal} {\bibinfo  {journal} {Phys. Rev. A}\ }\textbf {\bibinfo {volume} {87}},\ \bibinfo {pages} {062318} (\bibinfo {year} {2013})}\BibitemShut {NoStop}%
\bibitem [{\citenamefont {Gokhale}\ \emph {et~al.}(2019)\citenamefont {Gokhale}, \citenamefont {Baker}, \citenamefont {Duckering}, \citenamefont {Brown}, \citenamefont {Brown},\ and\ \citenamefont {Chong}}]{toffoli_decomposition_gokhale}%
  \BibitemOpen
  \bibfield  {author} {\bibinfo {author} {\bibfnamefont {P.}~\bibnamefont {Gokhale}}, \bibinfo {author} {\bibfnamefont {J.~M.}\ \bibnamefont {Baker}}, \bibinfo {author} {\bibfnamefont {C.}~\bibnamefont {Duckering}}, \bibinfo {author} {\bibfnamefont {N.~C.}\ \bibnamefont {Brown}}, \bibinfo {author} {\bibfnamefont {K.~R.}\ \bibnamefont {Brown}},\ and\ \bibinfo {author} {\bibfnamefont {F.~T.}\ \bibnamefont {Chong}},\ }\bibfield  {title} {\bibinfo {title} {Asymptotic improvements to quantum circuits via qutrits},\ }in\ \href@noop {} {\emph {\bibinfo {booktitle} {2019 ACM/IEEE 46th Annual International Symposium on Computer Architecture (ISCA)}}}\ (\bibinfo {year} {2019})\ pp.\ \bibinfo {pages} {554--566}\BibitemShut {NoStop}%
\bibitem [{\citenamefont {Chatterjee}\ \emph {et~al.}(2023{\natexlab{a}})\citenamefont {Chatterjee}, \citenamefont {Das}, \citenamefont {Bala}, \citenamefont {Saha}, \citenamefont {Chattopadhyay},\ and\ \citenamefont {Chakrabarti}}]{Chatterjee_2023}%
  \BibitemOpen
  \bibfield  {author} {\bibinfo {author} {\bibfnamefont {T.}~\bibnamefont {Chatterjee}}, \bibinfo {author} {\bibfnamefont {A.}~\bibnamefont {Das}}, \bibinfo {author} {\bibfnamefont {S.~K.}\ \bibnamefont {Bala}}, \bibinfo {author} {\bibfnamefont {A.}~\bibnamefont {Saha}}, \bibinfo {author} {\bibfnamefont {A.}~\bibnamefont {Chattopadhyay}},\ and\ \bibinfo {author} {\bibfnamefont {A.}~\bibnamefont {Chakrabarti}},\ }\bibfield  {title} {\bibinfo {title} {Qudiet: A classical simulation platform for qubit‐qudit hybrid quantum systems},\ }\href {https://doi.org/10.1049/qtc2.12058} {\bibfield  {journal} {\bibinfo  {journal} {IET Quantum Communication}\ }\textbf {\bibinfo {volume} {4}},\ \bibinfo {pages} {167–180} (\bibinfo {year} {2023}{\natexlab{a}})}\BibitemShut {NoStop}%
\bibitem [{\citenamefont {Srikara}\ and\ \citenamefont {Chandrashekar}(2020)}]{Srikara_2020}%
  \BibitemOpen
  \bibfield  {author} {\bibinfo {author} {\bibfnamefont {S.}~\bibnamefont {Srikara}}\ and\ \bibinfo {author} {\bibfnamefont {C.~M.}\ \bibnamefont {Chandrashekar}},\ }\bibfield  {title} {\bibinfo {title} {Quantum direct communication protocols using discrete-time quantum walk},\ }\bibfield  {journal} {\bibinfo  {journal} {Quantum Information Processing}\ }\textbf {\bibinfo {volume} {19}},\ \href {https://doi.org/10.1007/s11128-020-02793-4} {10.1007/s11128-020-02793-4} (\bibinfo {year} {2020})\BibitemShut {NoStop}%
\bibitem [{\citenamefont {Saha}\ \emph {et~al.}(2022{\natexlab{c}})\citenamefont {Saha}, \citenamefont {Majumdar}, \citenamefont {Saha}, \citenamefont {Chakrabarti},\ and\ \citenamefont {Sur-Kolay}}]{Saha_2022}%
  \BibitemOpen
  \bibfield  {author} {\bibinfo {author} {\bibfnamefont {A.}~\bibnamefont {Saha}}, \bibinfo {author} {\bibfnamefont {R.}~\bibnamefont {Majumdar}}, \bibinfo {author} {\bibfnamefont {D.}~\bibnamefont {Saha}}, \bibinfo {author} {\bibfnamefont {A.}~\bibnamefont {Chakrabarti}},\ and\ \bibinfo {author} {\bibfnamefont {S.}~\bibnamefont {Sur-Kolay}},\ }\bibfield  {title} {\bibinfo {title} {Faster search of clustered marked states with lackadaisical quantum walks},\ }\bibfield  {journal} {\bibinfo  {journal} {Quantum Information Processing}\ }\textbf {\bibinfo {volume} {21}},\ \href {https://doi.org/10.1007/s11128-022-03606-6} {10.1007/s11128-022-03606-6} (\bibinfo {year} {2022}{\natexlab{c}})\BibitemShut {NoStop}%
\bibitem [{\citenamefont {Gokhale}\ \emph {et~al.}(2020)\citenamefont {Gokhale}, \citenamefont {Baker}, \citenamefont {Duckering}, \citenamefont {Chong}, \citenamefont {Brown},\ and\ \citenamefont {Brown}}]{gokhale2020extending}%
  \BibitemOpen
  \bibfield  {author} {\bibinfo {author} {\bibfnamefont {P.}~\bibnamefont {Gokhale}}, \bibinfo {author} {\bibfnamefont {J.~M.}\ \bibnamefont {Baker}}, \bibinfo {author} {\bibfnamefont {C.}~\bibnamefont {Duckering}}, \bibinfo {author} {\bibfnamefont {F.~T.}\ \bibnamefont {Chong}}, \bibinfo {author} {\bibfnamefont {N.~C.}\ \bibnamefont {Brown}},\ and\ \bibinfo {author} {\bibfnamefont {K.~R.}\ \bibnamefont {Brown}},\ }\bibfield  {title} {\bibinfo {title} {Extending the frontier of quantum computers with qutrits},\ }\href@noop {} {\bibfield  {journal} {\bibinfo  {journal} {IEEE Micro}\ }\textbf {\bibinfo {volume} {40}},\ \bibinfo {pages} {64} (\bibinfo {year} {2020})}\BibitemShut {NoStop}%
\bibitem [{\citenamefont {Amy}\ \emph {et~al.}(2013)\citenamefont {Amy}, \citenamefont {Maslov}, \citenamefont {Mosca},\ and\ \citenamefont {Roetteler}}]{Amy_2013}%
  \BibitemOpen
  \bibfield  {author} {\bibinfo {author} {\bibfnamefont {M.}~\bibnamefont {Amy}}, \bibinfo {author} {\bibfnamefont {D.}~\bibnamefont {Maslov}}, \bibinfo {author} {\bibfnamefont {M.}~\bibnamefont {Mosca}},\ and\ \bibinfo {author} {\bibfnamefont {M.}~\bibnamefont {Roetteler}},\ }\bibfield  {title} {\bibinfo {title} {A meet-in-the-middle algorithm for fast synthesis of depth-optimal quantum circuits},\ }\href {https://doi.org/10.1109/tcad.2013.2244643} {\bibfield  {journal} {\bibinfo  {journal} {IEEE Transactions on Computer-Aided Design of Integrated Circuits and Systems}\ }\textbf {\bibinfo {volume} {32}},\ \bibinfo {pages} {818–830} (\bibinfo {year} {2013})}\BibitemShut {NoStop}%
\bibitem [{\citenamefont {Saha}\ and\ \citenamefont {Khanna}(2023)}]{sahaintermediatequdit}%
  \BibitemOpen
  \bibfield  {author} {\bibinfo {author} {\bibfnamefont {A.}~\bibnamefont {Saha}}\ and\ \bibinfo {author} {\bibfnamefont {O.}~\bibnamefont {Khanna}},\ }\bibfield  {title} {\bibinfo {title} {Intermediate-qudit assisted improved quantum algorithm for string matching with an advanced decomposition of fredkin gate},\ }\href@noop {} {\bibfield  {journal} {\bibinfo  {journal} {arXiv preprint arXiv:2304.03050}\ } (\bibinfo {year} {2023})}\BibitemShut {NoStop}%
\bibitem [{\citenamefont {Chatterjee}\ \emph {et~al.}(2023{\natexlab{b}})\citenamefont {Chatterjee}, \citenamefont {Das}, \citenamefont {Bala}, \citenamefont {Saha}, \citenamefont {Chattopadhyay},\ and\ \citenamefont {Chakrabarti}}]{sahaqudiet}%
  \BibitemOpen
  \bibfield  {author} {\bibinfo {author} {\bibfnamefont {T.}~\bibnamefont {Chatterjee}}, \bibinfo {author} {\bibfnamefont {A.}~\bibnamefont {Das}}, \bibinfo {author} {\bibfnamefont {S.~K.}\ \bibnamefont {Bala}}, \bibinfo {author} {\bibfnamefont {A.}~\bibnamefont {Saha}}, \bibinfo {author} {\bibfnamefont {A.}~\bibnamefont {Chattopadhyay}},\ and\ \bibinfo {author} {\bibfnamefont {A.}~\bibnamefont {Chakrabarti}},\ }\bibfield  {title} {\bibinfo {title} {Qudiet: A classical simulation platform for qubit-qudit hybrid quantum systems},\ }\href@noop {} {\bibfield  {journal} {\bibinfo  {journal} {IET Quantum Communication}\ } (\bibinfo {year} {2023}{\natexlab{b}})}\BibitemShut {NoStop}%
\bibitem [{\citenamefont {Di}\ and\ \citenamefont {Wei}(2012)}]{di5485elementary}%
  \BibitemOpen
  \bibfield  {author} {\bibinfo {author} {\bibfnamefont {Y.-M.}\ \bibnamefont {Di}}\ and\ \bibinfo {author} {\bibfnamefont {H.-R.}\ \bibnamefont {Wei}},\ }\bibfield  {title} {\bibinfo {title} {Elementary gates for ternary quantum logic circuit},\ }\href@noop {} {\bibfield  {journal} {\bibinfo  {journal} {arXiv preprint arXiv:1105.5485}\ } (\bibinfo {year} {2012})}\BibitemShut {NoStop}%
\bibitem [{\citenamefont {Lanyon}\ \emph {et~al.}(2009)\citenamefont {Lanyon}, \citenamefont {Barbieri}, \citenamefont {Almeida}, \citenamefont {Jennewein}, \citenamefont {Ralph}, \citenamefont {Resch}, \citenamefont {Pryde}, \citenamefont {O’brien}, \citenamefont {Gilchrist},\ and\ \citenamefont {White}}]{lanyon2009simplifying}%
  \BibitemOpen
  \bibfield  {author} {\bibinfo {author} {\bibfnamefont {B.~P.}\ \bibnamefont {Lanyon}}, \bibinfo {author} {\bibfnamefont {M.}~\bibnamefont {Barbieri}}, \bibinfo {author} {\bibfnamefont {M.~P.}\ \bibnamefont {Almeida}}, \bibinfo {author} {\bibfnamefont {T.}~\bibnamefont {Jennewein}}, \bibinfo {author} {\bibfnamefont {T.~C.}\ \bibnamefont {Ralph}}, \bibinfo {author} {\bibfnamefont {K.~J.}\ \bibnamefont {Resch}}, \bibinfo {author} {\bibfnamefont {G.~J.}\ \bibnamefont {Pryde}}, \bibinfo {author} {\bibfnamefont {J.~L.}\ \bibnamefont {O’brien}}, \bibinfo {author} {\bibfnamefont {A.}~\bibnamefont {Gilchrist}},\ and\ \bibinfo {author} {\bibfnamefont {A.~G.}\ \bibnamefont {White}},\ }\bibfield  {title} {\bibinfo {title} {Simplifying quantum logic using higher-dimensional hilbert spaces},\ }\href@noop {} {\bibfield  {journal} {\bibinfo  {journal} {Nature Physics}\ }\textbf {\bibinfo {volume} {5}},\ \bibinfo {pages} {134} (\bibinfo {year} {2009})}\BibitemShut {NoStop}%
\bibitem [{\citenamefont {Ralph}\ \emph {et~al.}(2007)\citenamefont {Ralph}, \citenamefont {Resch},\ and\ \citenamefont {Gilchrist}}]{ralph2007efficient}%
  \BibitemOpen
  \bibfield  {author} {\bibinfo {author} {\bibfnamefont {T.}~\bibnamefont {Ralph}}, \bibinfo {author} {\bibfnamefont {K.}~\bibnamefont {Resch}},\ and\ \bibinfo {author} {\bibfnamefont {A.}~\bibnamefont {Gilchrist}},\ }\bibfield  {title} {\bibinfo {title} {Efficient toffoli gates using qudits},\ }\href@noop {} {\bibfield  {journal} {\bibinfo  {journal} {Physical Review A}\ }\textbf {\bibinfo {volume} {75}},\ \bibinfo {pages} {022313} (\bibinfo {year} {2007})}\BibitemShut {NoStop}%
\bibitem [{\citenamefont {Shakeel}(2020)}]{Shakeel2020}%
  \BibitemOpen
  \bibfield  {author} {\bibinfo {author} {\bibfnamefont {A.}~\bibnamefont {Shakeel}},\ }\bibfield  {title} {\bibinfo {title} {Efficient and scalable quantum walk algorithms via the quantum fourier transform.}\ }\href {https://doi.org/10.1007/s11128-020-02834-y} {10.1007/s11128-020-02834-y} (\bibinfo {year} {2020})\BibitemShut {NoStop}%
\bibitem [{\citenamefont {Saha}\ \emph {et~al.}(2022{\natexlab{d}})\citenamefont {Saha}, \citenamefont {Majumdar}, \citenamefont {Saha}, \citenamefont {Chakrabarti},\ and\ \citenamefont {Sur-Kolay}}]{saha_n-qudit_toffoli_decomposition}%
  \BibitemOpen
  \bibfield  {author} {\bibinfo {author} {\bibfnamefont {A.}~\bibnamefont {Saha}}, \bibinfo {author} {\bibfnamefont {R.}~\bibnamefont {Majumdar}}, \bibinfo {author} {\bibfnamefont {D.}~\bibnamefont {Saha}}, \bibinfo {author} {\bibfnamefont {A.}~\bibnamefont {Chakrabarti}},\ and\ \bibinfo {author} {\bibfnamefont {S.}~\bibnamefont {Sur-Kolay}},\ }\bibfield  {title} {\bibinfo {title} {Asymptotically improved circuit for a $d$-ary grover's algorithm with advanced decomposition of the $n$-qudit toffoli gate},\ }\href {https://doi.org/10.1103/PhysRevA.105.062453} {\bibfield  {journal} {\bibinfo  {journal} {Phys. Rev. A}\ }\textbf {\bibinfo {volume} {105}},\ \bibinfo {pages} {062453} (\bibinfo {year} {2022}{\natexlab{d}})}\BibitemShut {NoStop}%
\bibitem [{\citenamefont {Bennett}\ \emph {et~al.}(1997)\citenamefont {Bennett}, \citenamefont {Bernstein}, \citenamefont {Brassard},\ and\ \citenamefont {Vazirani}}]{bennett1997strengths}%
  \BibitemOpen
  \bibfield  {author} {\bibinfo {author} {\bibfnamefont {C.~H.}\ \bibnamefont {Bennett}}, \bibinfo {author} {\bibfnamefont {E.}~\bibnamefont {Bernstein}}, \bibinfo {author} {\bibfnamefont {G.}~\bibnamefont {Brassard}},\ and\ \bibinfo {author} {\bibfnamefont {U.}~\bibnamefont {Vazirani}},\ }\bibfield  {title} {\bibinfo {title} {Strengths and weaknesses of quantum computing},\ }\href@noop {} {\bibfield  {journal} {\bibinfo  {journal} {SIAM journal on Computing}\ }\textbf {\bibinfo {volume} {26}},\ \bibinfo {pages} {1510} (\bibinfo {year} {1997})}\BibitemShut {NoStop}%
\bibitem [{\citenamefont {Benioff}(2000)}]{benioff2000space}%
  \BibitemOpen
  \bibfield  {author} {\bibinfo {author} {\bibfnamefont {P.}~\bibnamefont {Benioff}},\ }\bibfield  {title} {\bibinfo {title} {Space searches with a quantum robot},\ }\href@noop {} {\bibfield  {journal} {\bibinfo  {journal} {arXiv preprint quant-ph/0003006}\ } (\bibinfo {year} {2000})}\BibitemShut {NoStop}%
\bibitem [{\citenamefont {Aaronson}\ and\ \citenamefont {Ambainis}(2003)}]{aaronson2003quantum}%
  \BibitemOpen
  \bibfield  {author} {\bibinfo {author} {\bibfnamefont {S.}~\bibnamefont {Aaronson}}\ and\ \bibinfo {author} {\bibfnamefont {A.}~\bibnamefont {Ambainis}},\ }\bibfield  {title} {\bibinfo {title} {Quantum search of spatial regions},\ }in\ \href@noop {} {\emph {\bibinfo {booktitle} {44th Annual IEEE Symposium on Foundations of Computer Science, 2003. Proceedings.}}}\ (\bibinfo {organization} {IEEE},\ \bibinfo {year} {2003})\ pp.\ \bibinfo {pages} {200--209}\BibitemShut {NoStop}%
\bibitem [{\citenamefont {Childs}\ and\ \citenamefont {Goldstone}(2004)}]{childs2004spatial}%
  \BibitemOpen
  \bibfield  {author} {\bibinfo {author} {\bibfnamefont {A.~M.}\ \bibnamefont {Childs}}\ and\ \bibinfo {author} {\bibfnamefont {J.}~\bibnamefont {Goldstone}},\ }\bibfield  {title} {\bibinfo {title} {Spatial search by quantum walk},\ }\href@noop {} {\bibfield  {journal} {\bibinfo  {journal} {Physical Review A}\ }\textbf {\bibinfo {volume} {70}},\ \bibinfo {pages} {022314} (\bibinfo {year} {2004})}\BibitemShut {NoStop}%
\bibitem [{\citenamefont {Wong}(2018)}]{Wong_2018}%
  \BibitemOpen
  \bibfield  {author} {\bibinfo {author} {\bibfnamefont {T.~G.}\ \bibnamefont {Wong}},\ }\bibfield  {title} {\bibinfo {title} {Faster search by lackadaisical quantum walk},\ }\bibfield  {journal} {\bibinfo  {journal} {Quantum Information Processing}\ }\textbf {\bibinfo {volume} {17}},\ \href {https://doi.org/10.1007/s11128-018-1840-y} {10.1007/s11128-018-1840-y} (\bibinfo {year} {2018})\BibitemShut {NoStop}%
\end{thebibliography}%

\appendix


\section{Generalization of intermediate qudit approach}\label{appendix: intermediate}

    The intermediate qudit approach is generalized and further optimized in \cite{sahaintermediate} using two higher dimensions i.e., \ket{d+1} and \ket{d+2}. First, we discuss how it is generalized and then how it is optimized.
        \begin{enumerate}
            \item \textbf{Generalization:} The decomposition circuit for a d-ary 3-controlled Toffoli gate using the approach is shown in Figure \ref{subfiga:3-controlled-d-ary-toffoli-decomposition}. In this case, we are required to use only one higher-level state, \ket{d+1}. But to generalize the work of \cite{gokhale2020extending} for an $n$-controlled $d$-ary Toffoli gate, we need to use two higher-level states, \ket{d+1} and \ket{d+2}, as in Figure \ref{fig:d-ary-toffoli-generalized-decomposition-2qudit-gates}.
                \begin{figure}[!h]
                    \centering
                       \scalebox{.6}{
                        \begin{quantikz}
                            \lstick{\ket{q_0}}&\ctrl{1}&\\
                            \lstick{\ket{q_1}}&\ctrl{1}&\\
                            \lstick{\ket{q_2}}&\ctrl{1}&\\
                            \lstick{\ket{q_3}}&\gate{U_d}&
                        \end{quantikz}
                        $\equiv$
                        \begin{quantikz}
                            \lstick{\ket{q_0}}&\measure{d-1} \highlightCircuitTopLabel{3}{2}{blue}{Preparation}& & &\highlightCircuitTopLabel{3}{2}{red}{Correction}&\measure{d-1} &\\
                            \lstick{\ket{q_1}}&\gate{X_{d+1}^{+1}}\wire[u]{a}&\measure{d}& &\measure{d}&\gate{X_{d+1}^{-1}}\wire[u]{a}&\\
                            \lstick{\ket{q_2}}& &\gate{X_{d+1}^{+1}}\wire[u]{a}&\measure{d}\highlightCircuitBottomLabel{2}{1}{green}{Target Operation}&\gate{X_{d+1}^{-1}}\wire[u]{a}& & \\
                            \lstick{\ket{q_3}}& & &\gate{U_d}\wire[u]{a}&&&
                        \end{quantikz}}
                        \caption{3-controlled $d$-ary Toffoli decomposition using intermediate qudits of dimension $d$ and $d+1$. Here $U_d$ is used instead of $\oplus$ or $X_d$ to represent any unitary operation in general and not just \textit{Addition modulo $d$}.}
                        \label{subfiga:3-controlled-d-ary-toffoli-decomposition}
                    \end{figure}
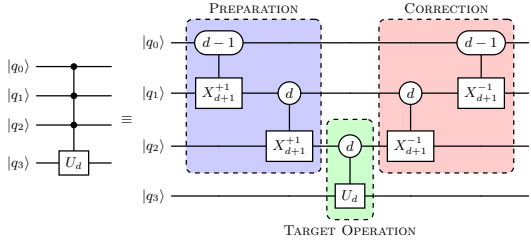

            \item \textbf{Optimization:} In Figure \ref{fig:d-ary-toffoli-generalized-decomposition-2qudit-gates}, we observe that  there are some redundant gates. Hence, an optimization has been used, which is proposed in \cite{sahaintermediate}. We now need to remove these redundant gates for a more optimized decomposition of the 7-controlled $d$-ary Toffoli gate, which is shown in Figure \ref{8-qudit}. 
            \begin{figure}[!h]
                \centering
                \scalebox{.5}{
                \begin{quantikz}[column sep=4pt, row sep={20pt,between origins}]
                    \lstick{\ket{q_0}} &\measure{d-1}\wire[d]{a}\highlightCircuitTopLabel{3}{3}{yellow}{i}& &\measure{d-1}\wire[d]{a}\highlightCircuitTopLabel{3}{1}{red}{Redundant}&&&&&&&&\measure{d-1}\wire[d]{a}\highlightCircuitTopLabel{3}{1}{red}{Redundant}& &\measure{d-1}\wire[d]{a}& \\
                    \lstick{\ket{q_1}} &\wire[d]{a}&\gate{X_{d+1}^{+1}}&\wire[d]{a}&\measure{d}\wire[d]{a}&&\measure{d}\wire[d]{a}\highlightCircuitTopLabel{5}{1}{red}{Redundant}&&\measure{d}\wire[d]{a}\highlightCircuitTopLabel{5}{1}{red}{Redundant}&&\measure{d}\wire[d]{a}&\wire[d]{a}&\gate{X_{d+1}^{-1}}&\wire[d]{a}& \\
                    \lstick{\ket{q_2}} &\gate{X_{d+1}^{+1}}&\measure{d}\wire[u]{a}&\gate{X_{d+1}^{-1}}&\wire[d]{a}& &\wire[d]{a}&&\wire[d]{a}&&\wire[d]{a}&\gate{X_{d+1}^{-1}}&\measure{d}\wire[u]{a}&\gate{X_{d+1}^{-1}}& \\
                    \lstick{\ket{q_3}} & & & &\wire[d]{a}&\gate{X_{d+1}^{+1}}&\wire[d]{a}&\measure{d}\wire[d]{a}&\wire[d]{a}&\gate{X_{d+1}^{-1}}&\wire[d]{a}&&&& \\
                    \lstick{\ket{q_4}} &\measure{d-1}\wire[d]{a}\highlightCircuitBottomLabel{3}{3}{yellow}{ii}& &\measure{d-1}\wire[d]{a}\highlightCircuitBottomLabel{3}{1}{red}{Redundant}&\wire[d]{a}&\wire[u]{a}&\wire[d]{a}&\wire[d]{a}&\wire[d]{a}&\wire[u]{a}&\wire[d]{a}&\measure{d-1}\wire[d]{a}\highlightCircuitBottomLabel{3}{1}{red}{Redundant}& &\measure{d-1}\wire[d]{a}& \\
                    \lstick{\ket{q_5}} &\wire[d]{a}&\gate{X_{d+1}^{+1}}&\wire[d]{a}&\gate{X_{d+2}^{+1}}&\measure{d+1}\wire[u]{a}&\gate{X_{d+2}^{-1}}&\wire[d]{a}&\gate{X_{d+2}^{+1}}&\measure{d+1}\wire[u]{a}&\gate{X_{d+2}^{-1}}&\wire[d]{a}&\gate{X_{d+1}^{+1}}&\wire[d]{a}& \\
                    \lstick{\ket{q_6}} &\gate{X_{d+1}^{+1}}&\measure{d}\wire[u]{a}&\gate{X_{d+1}^{-1}}&&&&\wire[d]{a}&&&&\gate{X_{d+1}^{-1}}&\measure{d}\wire[u]{a}&\gate{X_{d+1}^{-1}}& \\
                    \lstick{\ket{q_7}} & & & & & & &\gate{U_d}&&&&&&&
                \end{quantikz}}
                \caption{7-controlled Toffoli decomposition in $d$-dimensional systems. We have highlighted in yellow and labeled as $I$ and $II$ two 2-qudit gates to make it more visually understandable how the decomposition is done. By red color, redundant gates are highlighted which is further optimized.}
                \label{fig:d-ary-toffoli-generalized-decomposition-2qudit-gates}
            \end{figure}
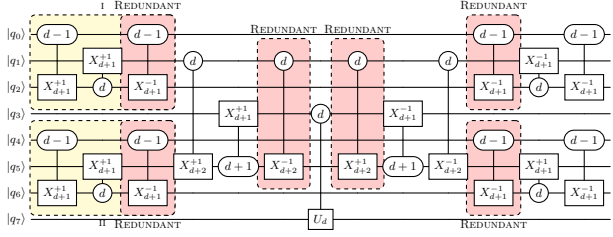

               \begin{figure}[!h]
            \centering
            \scalebox{.5}{
\begin{quantikz}
                \lstick{\ket{q_0}} &\measure{d-1}\wire[d]{a}\highlightCircuitTopLabel{7}{4}{blue}{Preparation}& & & & &\highlightCircuitTopLabel{7}{4}{red}{Correction}& & &\measure{d-1}\wire[d]{a}& \\
                \lstick{\ket{q_1}} &\wire[d]{a}&\gate{X^{+1}_{d+1}}&\measure{d}\wire[d]{a}& & & &\measure{d}\wire[d]{a}&\gate{X^{-1}_{d+1}}&\wire[d]{a}& \\
                \lstick{\ket{q_2}} &\gate{X^{+1}_{d+1}}&\measure{d}\wire[u]{a}&\wire[d]{a}& & & &\wire[d]{a}&\measure{d}\wire[u]{a}&\gate{X^{-1}_{d+1}}& \\
                \lstick{\ket{q_3}} & & &\wire[d]{a}&\gate{X^{+1}_{d+1}}\wire[d]{a}&\measure{d}\wire[d]{a}\highlightCircuitBottomLabel{5}{1}{green}{Target Operation}&\gate{X^{-1}_{d+1}}\wire[d]{a}&\wire[d]{a}&&& \\
                \lstick{\ket{q_4}} &\measure{d-1}\wire[d]{a}& &\wire[d]{a}&\wire[d]{a}&\wire[d]{a}&\wire[d]{a}&\wire[d]{a}& &\measure{d-1}\wire[d]{a}& \\
                \lstick{\ket{q_5}} &\wire[d]{a}&\gate{X^{+1}_{d+1}}&\gate{X^{+1}_{d+2}}&\measure{d+1}&\wire[d]{a}&\measure{d+1}&\gate{X^{-1}_{d+2}}&\gate{X^{-1}_{d+1}}&\wire[d]{a}& \\
                \lstick{\ket{q_6}} &\gate{X^{+1}_{d+1}}&\measure{d}\wire[u]{a}& & &\wire[d]{a}& & &\measure{d}\wire[u]{a}&\gate{X^{-1}_{d+1}}& \\
                \lstick{\ket{q_7}} & & & & &\gate{X}& & &&&
            \end{quantikz}}
            \caption{ Optimized 7-controlled generalized Toffoli decomposition using intermediate-qudit in $d$-dimensional systems.}
\label{8-qudit}
    \end{figure}
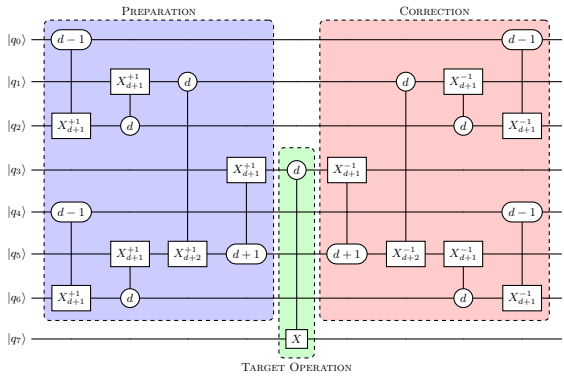
            
        \end{enumerate}

\nocite{*}


\end{document}